\documentclass[twocolumn,superscriptaddress,showpacs,aps,amsmath,amssymb,prb]{revtex4}
\usepackage{amsfonts}
\usepackage{mathrsfs}

\usepackage{bm}
\usepackage{graphicx}

\newcommand{\B}{\mbox{\tiny B}}

\newcommand{\R}{\mbox{\tiny R}}

\newcommand{\s}{\mbox{\tiny S}}

\newcommand{\nl}{\nonumber \\}

\newcommand{\Sec}[1]{Sec.\;\ref{#1}}

\newcommand{\be}{\begin{equation}}
\newcommand{\ee}{\end{equation}}
\newcommand{\bea}{\begin{eqnarray}}
\newcommand{\eea}{\end{eqnarray}}
\newcommand{\bsube}{\begin{subequations}}
\newcommand{\esube}{\end{subequations}}
\newcommand{\Eq}[1]{Eq.\,(\ref{#1})}

\newcommand{\dg}{\dagger}
\newcommand{\la}{\langle}
\newcommand{\ra}{\rangle}

\begin{document}

\title{A nonequilibrium theory for transient transport dynamics in nanostructures \\
via the Feynman-Vernon influence functional approach}

\author{Jinshuang Jin}
\affiliation{Department of Physics and Center for Quantum
Information Science, National Cheng Kung University, Tainan 70101,
Taiwan} \affiliation{ Department of Physics, Hangzhou Normal
University,
  Hangzhou 310036, China}
\author{Matisse Wei-Yuan Tu}
\affiliation{Department of Physics and Center for Quantum
Information Science, National Cheng Kung University, Tainan 70101,
Taiwan} \affiliation{Institute for Materials Science and Max
Bergmann Center of Biomaterials, Dresden University of Technology,
D-01069 Dresden, Germany.}
\author{Wei-Min Zhang} \email{wzhang@mail.ncku.edu.tw}
\affiliation{Department of Physics and Center for Quantum
Information Science, National Cheng Kung University, Tainan 70101,
Taiwan}
\author{YiJing Yan}
\affiliation{Department of Chemistry, Hong Kong University
   of Science and Technology, Kowloon, Hong Kong}


\begin{abstract}
In this paper, we develop a nonequilibrium theory for transient
electron transport dynamics in nanostructures based on the
Feynman-Vernon influence functional approach. We extend our previous
work on the exact master equation describing the non-Markovian
electron dynamics in the double dot [Phys.\,Rev.\,B78, 235311
(2008)] to the nanostructures in which the energy levels of the
central region, the couplings to the leads and the external biases
applied to leads are all time-dependent. We then derive
nonperturbatively the exact transient current in terms of the
reduced density matrix within the same framework. This provides an
exact non-linear response theory for quantum transport processes
with back-reaction effect from the contacts, including the
non-Markovian quantum relaxation and dephasing, being fully taken
into account. The nonequilibrium steady-state transport theory based
on the Schwinger-Keldysh nonequilibrium Green function technique can
be recovered as a long time limit.
For a simple application, we present the analytical and numerical
results of transient dynamics for a resonant tunneling nanoscale
device with a Lorentzian-type spectral density and ac bias
voltages, where the non-Markovian memory structure and non-linear
response to the time-dependent bias voltages in transport
processes are demonstrated.
\end{abstract}

 \maketitle

\section{introduction}

The investigation of quantum coherence dynamics far away from
equilibrium in nanoscale electronic devices has been attracting
much attentions in the past decade due to various applications in
nanotechnology and quantum information processing. Experimentally
it became possible not only to directly manufacture structures but
also to investigate their nonequilibrium quantum coherence
properties under well controlled
parameters.\cite{Gol98156,Cro98540,Lu03422,Hay03226084,Pet051280,Fuj06759,Han071217}
Theoretically, tremendous studies have been done focusing on the
time-independent steady-state to analyze the nonequilibrium
transport properties. However, an increasing number of work
employing various theoretical techniques (see
Refs.~[\onlinecite{Win938487,Jau945528,
Bru941076,Kon9616820,Sun004754,Zhu03224413,Li05205304,And05196801,
Lee06085325,Mac06085324,Jin08234703,Wei08195316,Sch08235110,Tu08235311,
Zed09045309,Zhe09164708,Hau98}]) has recently occurred addressing
the time-dependent electron dynamics far away from the
equilibrium. In fact, for practical applications, a full
understanding of nonequilibrium dynamics under external field
controls, i.e., the time evolution of electrons in nanoelectronics
devices from some initial preparation toward any specifically
designed state within an extremely short time, has been receiving
more and more attentions.

The prototypical nanostructure concerned in this paper is made by
a gate-defined region on a semiconductor containing a few of
discrete electronic states. Electrodes are implanted around this
central region to control the electrons across it. Also gates are
deposited to adjust the electronic states within the central area
as well as their couplings to the surrounding electrodes. Most of
the theoretical developments to nanoelectronics devices have been
focusing on the understanding and prediction of transport dynamics
in these devices. However, as a quantum device, in particular for
quantum information
processing,\cite{Hay03226084,Pet051280,Fuj06759,Han071217} the big
challenge is to understand and predict not only how fast or slow a
nanoelectronic device can turn on or off a current, but also how
reliable and efficient the device can manipulate quantum coherence
of the electron states through external bias and gate voltage
controls. This requires a nonequilibrium quantum device theory to
analyze the time-dependent quantum transport intimately entangling
with quantum coherence of electrons inside the device. It is the
purpose of this paper to attempt to establish a nonequilibrium
theory of transient dynamics for electron quantum states
accompanied with electron quantum transport in a nanoelectronic
device that can be feedbackly controlled through the non-linear
response to the external time-dependent bias and gate voltage
pulses.

Physically the nanostructure we are going to study is typically an
open quantum system in the sense that its central region has
exchanges of particles, energy and information with its
surroundings. From an open quantum system point of view, the
nonequilibrium electron dynamics is completely described by the
master equation through the physical observable $\langle O(t)
\rangle = {\rm tr}[ O \rho(t)]$, where $\rho(t)$ is the reduced
density matrix of the central system that can fully depict the
dynamics of electron quantum states and is determined by the master
equation. The transient electron transport should be in principle
solved from the reduced density matrix as well. However, it has been
struggled for many years without a very satisfactory answer in
finding the exact master equation for an arbitrary open quantum
system.\cite{Hua87} Most of the master equations used in the
literature are obtained using semiclassical approximation or
perturbation truncation, such as the semiclassical Boltzmann
equation\cite{Cer75,Smi89} or master equations under the Born-Markov
approximation.\cite{Car93,Bre02} However, for the nanostrucutres
with the extremely short length scales ($\sim$1 nm) and extremely
fast time scales ($\sim$1 fs), the semiclassical Boltzmann equation
or the master equation under the Born-Markov approximation are most
unlikely applicable.\cite{Hau98}

Literarily, there are two fundamental but equivalent methods in
dealing with modern nonequilibrium physics. They are the
Schwinger-Keldysh nonequilibrium Green function
technique\cite{Sch61407,Kad62,Kel651018} and the Feynman-Vernon
influence functional approach.\cite{Fey63118} In the practical
applications, both methods have their own advantages. The
Schwinger-Keldysh nonequilibrium Green function technique is a
standard method for systematic perturbation calculations to various
nonequilibrium physical processes in many-body
systems.\cite{Cho851,Ram86323} As it is evidenced the nonequilibrium
Green function technique provides an extremely powerful tool in the
study of the steady-state transport properties in
nanostructures.\cite{Mei913048,Mei922512,Dat95,Hau98} In the
steady-state limit where the initial state of the nanostructure is
irrelevant, the transport current can be expressed in terms of the
nonequilibrium Green functions in the frequency domain which largely
simplifies the transport problem. Without the time-dependence, the
physics of the nonequilibrium transport is essentially determined by
the density of states (the difference between the retarded and the
advanced Green functions). The nonequilibrium Green function
technique has been used to successfully describe a variety of new
phenomena, such as Kondo effect, Fano resonance and Coulomb blockade
effects in quantum dots. The extension of the nonequilibrium Green
function techniques to the time-dependent transport phenomena in
nanostructures has also been
developed\cite{Win938487,Jau945528,Sun004754,Zhu03224413,Hau98}
where the time-dependent nonequilibrium Green functions must be
maintained. Except for the wide band limit
(WBL),\cite{Jau945528,Hau98} the relevant Green function
calculations become rather complicated in order to taking into
account the different initial quantum state effect and the
non-Markovian memory structure.\cite{Mac06085324,Sch08235110}

On the other hand, the Feynman-Vernon influence functional
approach\cite{Fey63118} are mainly used to study dissipation
dynamics in quantum tunneling problems\cite{Leg871} and decoherence
problems in quantum measurement theory.\cite{Zuk03715} It is very
powerful to derive nonperturbatively the master equation for the
reduced density matrix of an open system by integrating out
completely the environmental degrees of freedom through the path
integral approach, where the non-Markovian memory structure is
manifested explicitly. In the early applications of the influence
functional approach, the master equation was derived for some
particular class of Ohmic (white-noise) environment for the quantum
Brownian motion (QBM) modeled as a central harmonic oscillator
linearly coupled to a set of harmonic oscillators simulating the
thermal bath.\cite{Cal83587,Haa852462} The exact master equation for
the QBM with a general spectral density (color-noise environments)
at arbitrary temperature that can fully address the non-Markovian
dynamics was the Hu-Paz-Zhang master equation.\cite{Hu922843}
Applications of the QBM exact master equation cover various topics,
such as quantum decoherence, quantum-to-classical transition,
quantum measurement theory, and quantum gravity and quantum
cosmology, etc.\cite{Wei99,Bre02,Hu08} Very recently, such an exact
master equation is further extended to the systems of two entangled
bosonic fields \cite{An07042127,Cho08011112,An090317} for the study
of non-Markovian entanglement dynamics.\cite{Yu04140404}
Nevertheless, using the influence functional approach to obtain the
exact master equation has been largely focused on the bosonic
(thermal) environments in the past half century.

Meanwhile, using master equation to study quantum electron transport
has also been attracting much attentions recently. With the
nonequilibrium Green function technique, the master equation can be
formally obtained in terms of real-time diagrammatic expansion up to
all the orders,\cite{Sch9418436,Kon961715,Kon9616820,Lee08224106}
although practically most of the master equation approach used in
quantum transport by far only take the perturbative theory up to the
second order.\cite{Leh02228305,Li05205304,Ped05195330,Wel06044712,
Har06235309,Cui06449} On the other hand, with the help of the
 influence functional approach, an alternative
formalism for the master equation for studying quantum transport and
electron dissipative dynamics has recently been developed in term of
the hierarchical expansion.\cite{Jin08234703} This new master
equation approach provides a very efficient tool for the numerical
study of the quantum transport properties, including the accurate
evaluations of the Coulomb blockage and Kondo transition
dynamics.\cite{Zhe08093016,Zhe09164708}

In this paper, we shall develop a nonequilibrium theory to describe
nonperturbatively the transient electron transport dynamics
intimately entangling with the electron quantum decoherence in
nanostructures. By extending the influence functional approach to
the fermionic reservoirs, we have recently derived an exact master
equation for a large class of nanostructures.\cite{Tu08235311} This
exact master equation is capable for the study of the full
non-Markovian decoherence dynamics of electrons in nanoelectronics
devices. In the present work, we shall extend the previous work to
the nanostructures in which the energy levels of the central region,
the couplings to the leads and the external biases applied to leads
are all time-dependent. We then derive the exact transient current
in terms of the reduced density matrix within the same framework.
The resulting transient current provides an exact non-linear
response theory that can be used to investigate the transient
quantum transport processes accompanied explicitly with the
non-Markovian quantum relaxation and dephasing dynamics. The
nonequilibrium steady-state transport theory based on the
nonequilibrium Green function technique that is widely used in the
literature can be recovered as a long time limit from the present
theory.

The remainder of this paper is organized as follows. In
\Sec{masterequation}, we outline the derivation of the exact master
equation we obtained recently,\cite{Tu08235311} with the extension
to the time-dependent Hamiltonian for a general nanostructure. In
\Sec{thcurr}, we derive in detail the exact transient current in
terms of the reduced density operator based on the Feynman-Vernon
influence functional approach. The resulting transient current
$I_\alpha(t)$ through the lead $\alpha$ is given by the general
formula, Eq.~(\ref{curr0}), or equivalently Eq.~(\ref{curr1}), which
is valid for arbitrary bias and gate voltage pulses with arbitrary
energy-dependent couplings between the central system and the leads.
The relation between the reduced density matrix and the transient
current is established through the master equation, see
Eq.~(\ref{DC}) where the superoperators ${\cal L}^{+}_\alpha(t)$ and
${\cal L}^{-}_\alpha(t)$ given explicitly by
Eq.~(\ref{current-sop1}) and (\ref{current-sop2}) encompass all the
back-reaction effects associated with the non-Markovian dynamics of
the central system interacting with the lead $\alpha$. The
steady-state current formula based on the nonequilibrium Green
function technique can be recovered as a long time limit, see
\Sec{thcurr}.E. Applications to transient dynamics with
Lorentzian-type spectral density and various time-dependent ac bias
voltages are demonstrated in \Sec{thdynam}. The summary and
prospective are given in \Sec{thsum}. In Appendix, a close
connection between the Feynman-Vernon influence functional approach
and the Schwinger-Keldysh nonequilibrium Green function technique
for quantum transport theory is established.

\section{Exact master equation for nanoscale electronic devices}
\label{masterequation}

In this section, we shall outline the exact master equation derived
recently by two of us for nanoscale electronic devices
\cite{Tu08235311} with the extension to the time-dependent energy
levels and couplings. We shall only list the necessary formula that
we will use later for the derivation of the transient current. For
more detailed derivation, please refer to
Ref.~[\onlinecite{Tu08235311}]. We begin with the Hamiltonian of a
prototypical mesoscopic system for electron transport,
\begin{align}
H(t)=&\sum_{ij}\epsilon_{ij}(t) a^\dg_i a_j + \sum_{\alpha k}
\epsilon_{\alpha k}(t) c^\dag_{\alpha k}c_{\alpha k} \notag \\
& +\sum_{i \alpha k}[V_{i\alpha k}(t)a^\dag_i c_{\alpha k}
+V^*_{i\alpha k}(t) c^\dag_{\alpha k}a_i]. \label{HH}
\end{align}
Here the first summation is the electron Hamiltonian $H_S$ for the
central system where the electron-electron interaction is not
considered. But as we have shown we can properly and explicitly
exclude the doubly occupied states and the resulting master equation
can be applied to the strong Coulomb repulsion (i.e.\,Coulomb
blockage) regime.\cite{Tu08235311} The second summation is the
Hamiltonian $\sum_\alpha H_{\alpha}$ describing the noninteracting
electron leads (the source and drain electrodes, $\cdots$) labeled
by the index $\alpha~(=L,R, \cdots$). The last term is the electron
tunneling Hamiltonian $H_T$ between the leads and the central
system. Since we will focus on the transient dynamics of quantum
transport in this paper, the bias voltages $V_\alpha$ applied to
leads are considered to be time-dependent. Thus the single electron
energy levels in the leads is changed as $\epsilon_{\alpha k}
\rightarrow \epsilon_{\alpha k}(t)=\epsilon_{\alpha
k}+eV_\alpha(t)$. Meanwhile, the energy levels $\epsilon_{ij}$ of
the central system and the couplings $V_{i\alpha k}$ between the
central system and the leads are controllable through the gate
voltages and external fields so that they can be, in general, also
time-dependent: $\epsilon_{ij} \rightarrow \epsilon_{ij}(t) =
\epsilon_{ij} + \Delta_{ij}(t), V_{i\alpha k} \rightarrow V_{i\alpha
k}(t)$. Throughout this work, we set $\hbar=1$, except for the
transient current where we will put $\hbar$ back from its
definition.

The master equation for the nonequilibrium electron dynamics of
the central system is derived based on the Feynman-Vernon
influence functional approach.\cite{Fey63118} As it is well
known,\cite{Hua87} the nonequilibrium electron dynamics of an open
system is completely determined by the so-called reduced density
matrix. The reduced density matrix is defined from the density
matrix of the total system (the central system plus the leads, and
the leads are treated as the reservoirs to the central system) by
tracing over entirely the environmental degrees of freedom:
$\rho(t) \equiv {\rm tr}_{\R} \rho_{\rm tot}(t)$, where the total
density matrix is formally given by $\rho_{\rm tot}(t)=U(t,t_0)
\rho_{\rm tot}(t_0) U^\dag(t,t_0)$ with the evolution operation
$U(t,t_0)=T \exp\{-i\int^t_{t_0}H(\tau)d\tau\}$, and $T$ is the
time-ordering operator. As usual, we assume that the central
system is uncorrelated with the reservoirs before the tunneling
couplings are turned on:\cite{Leg871} $\rho_{\rm
tot}(t_0)=\rho(t_0) \otimes \rho_{R}(t_0)$, and the reservoirs are
initially in the equilibrium state: $\rho_R(t_0)={1\over
Z}e^{-\sum_\alpha \beta_\alpha (H_\alpha-\mu_\alpha N_\alpha)}$
where $\beta_\alpha=1/(k_{\B}T_\alpha)$ is the initial inverse
temperature and $N_\alpha=\sum_k c^\dag_{\alpha k}c_{\alpha k}$
the particle number operator for the lead $\alpha$. Then the
reduced density matrix at an arbitrary later time $t$ can be
expressed in terms of the coherent state
representation\cite{Fad80,Zha90867} as
\begin{align}
\langle \bm \xi_f|\rho(t)|\bm \xi'_f\rangle=\int d\mu(\bm \xi_0)&
d\mu(\bm \xi'_0) \langle \bm \xi_0|\rho(t_0)|\bm \xi'_0\rangle \nonumber \\
& \times {\cal J}(\bar{\bm \xi}_f,\bm \xi'_f,t|\bm \xi_0,\bar{\bm
\xi}'_0,t_0), \label{rdm}
\end{align}
with $\bm \xi=(\xi_1, \xi_2, \cdots)$ and $\bar{\bm \xi}=(\xi^*_1,
\xi^*_2, \cdots)$ being the Grassmannian numbers and their complex
conjugate defined through the fermion coherent states: $a_i|\bm
\xi\rangle = \xi_i |\bm \xi \rangle$ and $\langle \bm \xi |a^\dag_i
= \langle \bm \xi | \xi^*_i $. The propagating function in
Eq.~(\ref{rdm}) is given in terms of Grassmannian number path
integrals,
\begin{align}
{\cal J}(\bar{\bm \xi}_f,&\bm \xi'_f,t|\bm \xi_0,\bar{\bm \xi}'_0,t_0)=\nonumber\\
&\int\mathcal{D}[\bar{\bm \xi} \bm \xi;\bar{\bm \xi}' \bm \xi']e^{i
(S_{c}[\bar{\bm \xi},~\bm \xi]-S^*_{c}[\bar{\bm \xi}', ~\bm
\xi'])}\mathcal{F} [\bar{\bm \xi} \bm \xi;\bar{\bm \xi}' \bm \xi'] ,
\label{ppg}
\end{align}
where $S_{c}[\bar{\bm \xi},\bm \xi]$ is the action of the central
system in the fermion coherent state representation, and
$\mathcal{F}[\bar{\bm \xi} \bm \xi;\bar{\bm \xi}' \bm \xi']$ is the
influence functional obtained after integrating out all the
environmental (reservoirs) degrees of freedom:
\begin{widetext}
\begin{align}
\mathcal{F}_{}[\bar{\bm \xi} \bm \xi;\bar{\bm \xi}' \bm \xi']=\exp
\Big\{& - \sum_{\alpha ij} \int_{t_0}^{t}d\tau
\int_{t_0}^{\tau}d\tau'\Big( \bm g_{\alpha ij}
(\tau,\tau')\xi^*_{i}(\tau)\xi_{j}(\tau') + \bm g^{*}_{\alpha
ij}(\tau,\tau')\xi'^*_{i}(\tau')\xi'_{j}(\tau)\Big) \nonumber\\
& +\sum_{\alpha ij} \int_{t_0}^{t}d\tau\int_{t_0}^{t}d\tau'\Big( \bm
g_{\alpha ij}(\tau,\tau') \xi^*_{i}(\tau) \xi'_{j}(\tau') -
\bm g_{\alpha ij}^\beta(\tau,\tau') \big(\xi^*_{i}(\tau)
+\xi'^*_{i}(\tau)\big)\big(\xi_{j}(\tau')
+\xi'_{j}(\tau')\big) \Big)\Big\}. \label{IF}
\end{align}
\end{widetext}
The two-time correlations in Eq.~(\ref{IF}) are defined by
\begin{subequations}
\label{ggbeta}
 \begin{align}
  \label{gvt}
 \bm g_{\alpha ij}(\tau,\tau')
 &=\sum_{k} V_{i\alpha k}(\tau)V^*_{j \alpha k}(\tau')
   e^{-i\int^{\tau}_{\tau'}d\tau_1 \epsilon_{\alpha k}(\tau_1)},
 \\
 \bm g^{\beta }_{\alpha ij}(\tau,\tau')
 &=\sum_{k} V_{i\alpha k}(\tau)V^*_{j \alpha k}(\tau')
 f_\alpha(\epsilon_{\alpha k})
   e^{-i\int^{\tau}_{\tau'}d\tau_1 \epsilon_{\alpha k}(\tau_1)}
 \label{gbetavt}
 \end{align}
 \end{subequations}
which depict all the time-correlations of electrons in the leads
through the interactions with the central system, and $f_\alpha
(\epsilon_{\alpha k})= 1/(e^{\beta_\alpha (\epsilon_{\alpha
k}-\mu_\alpha)} -1)$ is the Fermi distribution function of the
lead $\alpha$ at the initial time $t_0$.

This influence functional takes fully into account the back-reaction
effects of the reservoirs to the central system. It modifies the
original action of the system into an effective one,
$e^{{i\over\hbar}(S_{c}[\bar{\bm \xi},\bm \xi]- S^*_{c}[\bar{\bm
\xi}',\bm \xi'])}\mathcal{F}[\bar{\bm \xi} \bm \xi;\bar{\bm \xi}'
\bm \xi'] =e^{{i\over\hbar}S_{\rm eff}[\bar{\bm \xi} \bm
\xi;\bar{\bm \xi}' \bm \xi'] }$ which dramatically changes the
dynamics of the central system. The detailed change is manifested
through the generating function of Eq.~(\ref{ppg}) by carrying out
the path integral with respect to the effective action $S_{\rm
eff}[\bar{\bm \xi} \bm \xi;\bar{\bm \xi}' \bm \xi']$. While the path
integral $\mathcal{D}[\bar{\bm \xi} \bm \xi;\bar{\bm \xi}' \bm
\xi']$ integrates over all the forward paths $\bar{\bm \xi}(\tau),
\bm \xi(\tau)$ and the backward paths $\bar{\bm \xi}'(\tau), \bm
\xi'(\tau)$ in the Grassmannian space bounded by $\bar{\bm
\xi}(t)=\bar{\bm \xi}_{f}, \bm \xi(t_0)=\bm \xi_0$ and $\bar{\bm
\xi}'(t_0)=\bar{\bm \xi}'_0, \bm \xi'(t)=\bm \xi'_{f}$,
respectively. Since $S_{\rm eff}[\bar{\bm \xi} \bm \xi;\bar{\bm
\xi}' \bm \xi']$ is only a quadratic function in terms of the
integral variables, the path integrals in Eq.~(\ref{ppg}) can be
reduced to Gaussian integrals so that we can use the stationary path
method to exactly carry out these path integrals.\cite{Fey65} The
resulting generating function is:\cite{Tu08235311}
\begin{align}
{\cal J}(\bar{\bm \xi}_{f}, & \bm \xi'_{f},t| \bm \xi_{0},\bar{\bm
\xi}'_{0},t_0)= \frac{1}{{\rm det}[\bm w^\beta(t)]} \exp
\Big\{\bar{\bm \xi}_f \bm J_1 (t)\bm \xi_0  \nonumber
\\ & + \bar{\bm \xi}_f \bm J_2(t)\bm \xi'_f + \bar{\bm \xi}'_0 \bm J_3(t) \bm \xi_0
+ \bar{\bm \xi}'_0 \bm J^\dagger_1(t) \bm \xi'_f \Big\} ,
\label{ppg1}
\end{align}
in which the time-dependent coefficients are given explicitly as:
$\bm J_1(t)=\bm w^\beta(t) \bm u(t)$, $\bm J_2(t)=\bm
w^\beta(t)-\bm I$, and $\bm J_3(t)=\bm u^{\dag}(t)\bm
w^\beta(t)\bm u(t)-\bm I$, with $\bm w^\beta(t)=[I-\bm
v^\beta(t)]^{-1}$. Here, we have also expressed the stationary
paths in terms of $N\times N$ matrix variables $\bm u(\tau)$, $\bm
v^\beta(\tau)$ and $\bar{\bm u}(\tau)$ for $t_0 \le \tau \le t$,
where $N$ is the total number of single particle energy levels in
the central region, and the index $\beta$ implies that the
corresponding function is temperature dependent. They satisfy the
following dissipation-fluctuation integrodifferential equations
(i.e. the stationary path equations of motion):
\bsube\label{uv-eq}
\begin{align}
\dot{\bm u}(\tau)+i\bm \epsilon(\tau) {\bm u}(\tau) + & \sum_\alpha
\int_{t_0}^{\tau } d\tau' \bm g_\alpha (\tau,\tau') {\bm
u}(\tau')=0, \label{ut-eq} \\
\dot{\bar{\bm u}}(\tau)+i \bm \epsilon (\tau)\bar{\bm u}(\tau) - &
\sum_\alpha  \int^{t}_{\tau } d\tau' \bm g_\alpha (\tau,\tau')
\bar{\bm u}(\tau')=0, \label{ubt-eq} \\
\dot{\bm  v}^\beta(\tau)+i\bm \epsilon (\tau)\bm  v^\beta(\tau) & +
\sum_\alpha \int_{t_0}^{\tau } d\tau' \bm g_\alpha (\tau, \tau') \bm
v^\beta(\tau') \notag \\ & = \sum_\alpha \int_{t_0}^{t }d\tau'
      \bm g^\beta_\alpha (\tau,\tau')\bar{\bm u}(\tau'),
\label{vt-eq}
\end{align}
\esube with the boundary conditions $\bm u(t_0)=\bm 1$, $\bar{\bm
u}(t)=\bm 1$ and $\bm v^\beta(t_0)=0$, where $\bm g_\alpha(\tau,
\tau')$ and $\bm g^\beta_\alpha(\tau, \tau')$ are the non-local
two-time correlation matrix functions, whose matrix elements are
given by Eq.~(\ref{ggbeta}). As we will show in Appendix the matrix
variables $\bm u(\tau)$,  $\bar{\bm u}(\tau)$ and $\bm
v^\beta(\tau)$ correspond respectively to the retarded, advanced and
lesser Green functions, and  $\bm g_\alpha(\tau, \tau')$ and $\bm
g^\beta_\alpha(\tau, \tau')$ are the retarded and lesser
self-energies in the nonequilibrium Green function technique. For
the most general cases where all the parameters in the Hamiltonian
(\ref{HH}) are time-dependent, the time translational invariance is
broken down. Then $\bm u(\tau)$ and $\bar{\bm u}(\tau)$ are usually
independent except for the end-points where we have $\bar{\bm
u}(t_0)=\bm u^\dag(t)$. If all the parameters in the Hamiltonian of
Eq.~(\ref{HH}) are the time-independent, $\bm u(\tau)$ will become
only a function of $\tau-t_0$ and $\bar{\bm u}(\tau) =\bm
u^\dag(t-\tau+t_0)$, as we have shown in
Ref.~[\onlinecite{Tu08235311}].

Taking the time derivative of the reduced density matrix with the
solution of the propagating function (\ref{ppg1}), together with the
$D$-algebra of fermion creation and annihilation operators in the
fermion coherent state representation, we arrive at the final form
of the exact master equation we obtained previously
\cite{Tu08235311}
\begin{align}
&\frac{d \rho(t)}{dt}=-i[H'_S(t),\rho(t)] \nonumber \\
& ~~~+ \sum_{ij}\Big\{\bm \gamma_{ij}(t)(2a_{j}\rho(t) a^{\dag}_{i}-
a^{\dag}_{i}a_{j}\rho(t)-\rho(t) a^{\dag}_{i}a_{j}) \nonumber\\&
~~~+\bm \gamma^\beta_{ij}(t)(a_{j}\rho(t) a^{\dag}_{i} -
a^{\dag}_{i}\rho(t) a_{j}- a^{\dag}_{i}a_{j}\rho(t) +\rho(t)
a_{j}a^{\dag}_{i}) \Big\}. \label{emaster}
\end{align}
The first term (the commutator) in the master equation accounts
for the renormalized effect (including the time-dependent shifts
of the energy levels and the changes of the transition amplitudes
between them) of the central system due to the interaction with
the leads. The resulting renormalized Hamiltonian is
$H'_S(t)=\sum_{ij}\bm \epsilon'_{ij}(t)a^\dg_i a_j$. While the
rest terms in the master equation describe the dissipation and
noise effects (which results in a non-unitary evolution of the
central system) induced also by the interaction with the leads.
All the time-dependent coefficients in Eq.~(\ref{emaster}) are
determined explicitly by $\bm u(t), \bm u^\dag(t)$ and $\bm
v^\beta(t)$ as follows:
\begin{subequations}
\label{td-coe}
\begin{align}
\bm \epsilon'_{ij}(t) &= {i\over 2}[\dot{\bm u}\bm u^{-1}- (\bm
u^\dag)^{-1}\dot{\bm u}^\dag]_{ij} \notag \\
& = \bm \epsilon_{ij}(t) -\frac{i}{2} \sum_\alpha
[\bm \kappa_\alpha(t) - \bm \kappa^\dag_\alpha(t)]_{ij} ,
\\
\bm \gamma_{ij}(t) &= -{1\over 2}[\dot{\bm u}\bm u^{-1}+ (\bm
u^\dag)^{-1}\dot{\bm u}^\dag]_{ij} \notag \\
& = \frac{1}{2} \sum_\alpha [\bm \kappa_\alpha(t) +
\bm \kappa^\dag_\alpha (t)]_{ij}, \label{gamma1}
\\
\bm \gamma^\beta_{ij}(t)&=[\dot{\bm u}\bm u^{-1}\bm v^\beta +\bm
v^\beta(\bm
u^\dag)^{-1}\dot{\bm u}^\dag-\dot{\bm v}^\beta]_{ij} \notag \\
& = \sum_\alpha [\bm \lambda^\beta_\alpha(t) + \bm \lambda^{\beta
\dag}_\alpha(t)]_{ij}. \label{Gam-beta}
\end{align}
\end{subequations}
Here we have used the relations obtained from Eq.~(\ref{uv-eq}),
\bsube \label{uu-sol}
\begin{align}
\bm \kappa_\alpha(t)= & \int_{t_0}^{t}d\tau \bm
g_\alpha(t,\tau)\bm u(\tau) [\bm u(t)]^{-1} ,  \label{uu-2} \\
\bm \lambda^\beta_\alpha(t) = & \int_{t_0}^{t} d\tau \big\{ \bm
g_{\alpha}(t,\tau) \bm v^\beta(\tau) -\bm
g^{\beta}_{\alpha}(t,\tau)\bar{\bm u}(\tau) \big\} \notag \\ & -\bm
\kappa_\alpha(t) \bm v^\beta(t) . \label{uu-3}
\end{align}
\esube

As it is well known, the time-dependent coefficients in the master
equation describe the non-Markovian dynamics due to the
interaction between the central system and the leads. The
time-dependence of these coefficients is fully determined by the
functions $\bm u(\tau), \bar{\bm u}(\tau)$ and $\bm v^\beta(\tau)$
which are governed by the integrodifferential equations
(\ref{uv-eq}) where the time integrals involve the non-local time
correlation functions of the reservoirs, $\bm
g_\alpha(\tau,\tau')$ and $\bm g^\beta_\alpha(\tau,\tau')$. These
non-local time correlation functions characterize all the
non-Markovian memory structures of the central system interacting
with its environment through the coupling Hamiltonian $H_T$. By
solving Eq.~(\ref{uv-eq}), one can completely describe the quantum
decoherence dynamics of electrons in the central region due to the
entanglement between the central system and the leads. Thus, the
master equation determines indeed the exact nonequilibrium
dynamics of the system. It may be also worth mentioning that our
master equation does not need to be in Lindblad form to preserve
positivity since it is exact so that the dampening coefficients
are time-dependent and the positivity is guaranteed. In the next
section, we will derive the transient current passing through the
central system within the same framework. The result will allow us
to explore the transient dynamics of the nonequilibrium quantum
transport accompanied with the non-Markovian decoherence
phenomena, which receives more and more attentions recently due to
rapid developments in nano-technology, spintronics and quantum
information processing, etc.\cite{Fuj06759,Han071217}

\section{Exact transient current}
\label{thcurr}
\subsection{Influence functional approach}
The transient current from the $\alpha$-lead tunneling through
$\alpha$-junction into the central region is defined in the
Heisenberg picture as
\begin{align} \label{curr_def} I_{\alpha}(t)=-e\, \la
\frac{d}{dt} {\hat N}_{\alpha}(t) \ra =i\frac{e}{\hbar}\,\big\la[\hat
N_{\alpha}(t),H(t)]\big\ra.
\end{align}
Here, $e$ is the electron charge,
$\la O(t) \ra \equiv {\rm Tr} [O(t) \rho^H_{\rm tot}]$ with ${\rm
tr} \equiv {\rm tr}_{\s}{\rm tr}_{\R}$, $\hat N_\alpha(t)=\sum_{k}
c^\dg_{\alpha k}(t)c_{\alpha k}(t)$, and $\rho^H_{\rm tot}$ is the
total density matrix in the Heisenberg picture. By explicitly
calculating the above commutation relation with the Hamiltonian of
Eq.~(\ref{HH}) and then transforming it into the
Sch\"{o}rdinger picture, we have
\begin{align}
I_\alpha(t)&=i \frac{e}{\hbar} \,{\rm tr}_{\s} \sum_i \big[A^\dag_{\alpha i}(t)a_i
 -a^\dg_i A_{\alpha i}(t)\big],        \label{tcd}
\end{align}
where the operators $A_{\alpha i}(t)  = {\rm tr}_{\R} \big[\sum_k
V_{i \alpha k}(t) c_{\alpha k} \rho_{\rm tot}(t)\big]$ and
$A^\dag_{\alpha i}(t) = [A_{\alpha i}(t)]^\dag$. The time-dependence
of these two operators comes from the non-Markovian memory dynamics
by tracing over the environmental degrees of freedom of the total
density matrix $\rho_{\rm tot}(t)$ in the Schr\"{o}dinger picture.
These two operators are indeed the effective (dressed) electronic
creation and annihilation operators acting on the reduced density
matrix of the central system, as a result of tracing over the
reservoir degrees of freedom for the corresponding operators acting
on the lead $\alpha$.

Following the procedure of obtaining the reduced density matrix
through the influence functional approach,\cite{Tu08235311} we can
write
\begin{align}
\langle\bm \xi_f|A_{\alpha i}(t)|\bm \xi'_f\rangle=\int & d\mu(\bm
\xi_0) d\mu(\bm \xi'_0) \langle\bm \xi_0|\rho(t_0)|\bm \xi'_0\rangle
\nonumber
\\ & \times {\cal J}^A_{\alpha i}(\bar{\bm \xi}_{f}, \bm \xi'_{f},t|
\bm \xi_{0},\bar{\bm \xi}'_{0},t_0), \label{effAO}
\end{align}
where the operator-associated propagating function is defined as
\begin{align}
{\cal J}^A_{\alpha i}(\bar{\bm \xi}_{f}, & \bm \xi'_{f},t| \bm
\xi_{0},\bar{\bm \xi}'_{0},t_0)=\nonumber\\
&\int\mathcal{D}[\bar{\bm \xi} \bm \xi;\bar{\bm \xi}' \bm \xi']e^{i
(S_{c}[\bar{\bm \xi},~\bm \xi]-S^*_{c}[\bar{\bm \xi}',~\bm
\xi'])}\mathcal{F}^A_{\alpha i} [\bar{\bm \xi} \bm \xi;\bar{\bm
\xi}' \bm \xi']. \label{AO-ppg}
\end{align}
Similar to the calculation of the influence functional in
\Eq{AO-ppg}, the operator-associated influence functional
$\mathcal{F}^A_{\alpha i} [\bar{\bm \xi} \bm \xi;\bar{\bm \xi}' \bm
\xi']$ can be calculated in the same way with the result,
\begin{align}\label{AIF}
  {\cal F}^A_{\alpha i}[\bar{\bm \xi} \bm \xi;\bar{\bm \xi}' \bm
\xi']= -i{\cal A}_{\alpha i}[\bm \xi,\bm \xi'] {\cal F}[\bar{\bm
\xi} \bm \xi;\bar{\bm \xi}' \bm \xi'],
\end{align}
where the functional expression of the effective electron
annihilation operator is obtained as
\begin{align}\label{calA}
{\cal \bm A}_{\alpha}[\bm \xi, \bm \xi']=
  \int_{t_0}^{t}\!d\tau \Big\{ \bm g_{\alpha }(t,\tau) \bm {\xi}(\tau)
- \bm g^{\beta}_{\alpha }(t,\tau) \big[\bm {\xi}(\tau)
    +\bm {\xi}'(\tau) \big] \Big\},
\end{align}
and the same influence functional ${\cal F}[\bar{\bm \xi} \bm
\xi;\bar{\bm \xi}' \bm \xi']$ is given by Eq.~(\ref{IF}). It should
be pointed out that the factorability of the operator-associated
influence functional is due to the fact that the environmental path
integral is exactly computable.

Since the path integrals in Eq.~(\ref{AO-ppg}) can also be reduced
to Gaussian integrals, similar to the derivation of the master
equation for the reduced density matrix (\ref{emaster}), we use
again the stationary phase method to exactly carry out the path
integrals of Eq.~(\ref{AO-ppg}). The result is:
\begin{align}
{\cal J}^A_{\alpha i}(\bar{\bm \xi}_{f}, &\bm \xi'_{f},t| \bm
\xi_{0},\bar{\bm \xi}'_{0},t_0)=\nonumber\\
&-i{\cal A}_{\alpha i}(\bm \xi'_f,\bm \xi_0, t) {\cal J} (\bar{\bm
\xi}_{f}, \bm \xi'_{f},t| \bm \xi_{0},\bar{\bm \xi}'_{0},t_0),
\label{A-ppg}
\end{align}
where
${\cal A}_{\alpha i}(\bm \xi'_f,\bm \xi_0, t) =\sum_j[\bm y_{\alpha
ij}(t)\bm \xi_{0j} + \bm z_{\alpha ij}(t) \bm \xi'_{f j}]$
with $\bm y_\alpha(t)=\int^t_{t_0}d\tau \bm g_\alpha(t,\tau)\bm
u(\tau)+\bm z_\alpha(t)\bm u(t)$, and $\bm z_\alpha(t)=[\bm
\lambda^\beta_\alpha(t) +\bm \kappa_\alpha(t) \bm v^\beta(t)][1-\bm
v^\beta(t)]^{-1}$. The propagating function ${\cal J} (\bar{\bm
\xi}_{f}, \bm \xi'_{f},t| \bm \xi_{0},\bar{\bm \xi}'_{0},t_0)$ is
given by Eq.~(\ref{ppg1}) in the last section. Substituting
Eq.~(\ref{A-ppg}) into Eq.~(\ref{effAO}), and using the identity,
\begin{align}
\bm \xi_0 {\cal J} = \bm u^{-1}(t)\Big\{[1-\bm
v^\beta(t)]\frac{\partial}{\partial \bm \xi_f} - \bm v^\beta(t)\bm
\xi'_f\Big\} {\cal J},   \notag
\end{align}
together with the $D$-algebra for the fermion creation and
annihilation operators in the fermion coherent state
representation, we obtain the effective electronic annihilation
operator after tracing over completely the environmental degrees
of freedom:
\begin{align}
A_{\alpha i}(t)= -i \sum_j  \Big\{& \bm \lambda^\beta_{\alpha ij}(t)
[a_j\rho(t)+\rho(t)a_j]+ \bm \kappa_{\alpha ij}(t) a_j \rho(t)
\Big\}, \label{effao}
\end{align}
where the coefficient matrices $\bm \lambda^\beta_\alpha(t)$ and
$\bm \kappa_\alpha(t)$ are the same as these appeared in the master
equation (\ref{emaster}) and are explicitly given by
Eq.~(\ref{uu-sol}), $\rho(t)$ is just the reduced density matrix
determined by the master equation.

Accordingly, substituting the above result into Eq.~(\ref{tcd}), the transient
current can be directly calculated. The resulting transient current is
rather simple:
\begin{align}
I_\alpha(t) = - & \frac{e}{\hbar}\,{\rm Tr}\big\{\bm
\lambda^\beta_\alpha(t) + \bm
 \kappa_\alpha(t)\bm \rho^{(1)}(t) + {\rm H.c.} \big\} \notag \\
 =- & \frac{2e}{\hbar}\, {\rm Re}\int_{t_0}^{t}\!d\tau \,
{\rm Tr}\Big\{ \bm g_{\alpha}(t,\tau)\bm v^\beta(\tau) -\bm
   g^{\beta}_{\alpha}(t,\tau)\bar{\bm u}(\tau) ~~\notag \\
 & + \bm g_\alpha(t,\tau)\bm u(\tau)\bm u^{-1}(t)
   [\bm \rho^{(1)}(t)-\bm v^\beta(t)] \Big\}, \label{curr0}
 \end{align}
where the notation ${\rm Tr}$ is the trace over the energy level basis of the
central system. $\bm \rho^{(1)}_{ij}(t) \equiv {\rm tr}_s [a^\dag_ja_i
\rho(t))]$ is the single particle reduced density matrix. While
the matrix elements of $\bm
g_\alpha(\tau, \tau')$ and $\bm g^\beta_\alpha(\tau, \tau')$ are
 defined by Eq.~(\ref{ggbeta}), and $\bm u(\tau)$, $\bm v^\beta(\tau)$ and
$\bar{\bm u}(\tau)$ are determined nonperturbatively by
Eq.~(\ref{uv-eq}), as shown in the last section. Thus the transient
current $I_\alpha (t)$ that flows from the lead $\alpha$ is
completely determined within the framework of the master equation
for the reduced density matrix. As we shall show later, from the
above result we can easily reproduce the steady-state current at
$t\rightarrow \infty$ in terms of the non-equilibrium Green
functions and the Landauer-B\"{u}ttiker formula.

\subsection{Relation between the transient current
 and the reduced density matrix}
In fact, the transient current defined by Eq.~(\ref{tcd}) can be
written as a trace over the transient current operator: $I_\alpha (t)
= \frac{e}{\hbar}\, {\rm tr}_{\rm s}[\hat{I}_{\alpha}(t)]$, where
the current operator is obtained directly from Eq.~(\ref{effao}):
\begin{align}
\hat{I}_\alpha(t)= & - \sum_{ij} \Big\{ \bm \lambda^\beta_{\alpha
ij}(t)
[a^\dag_i a_j \rho(t) + a^\dag_i \rho(t)a_j] \notag \\
&+ \bm \kappa_{\alpha ij}(t) a^\dag_i a_j \rho(t) + {\rm H.c.}
\Big\}\equiv {\cal L}^{+}_\alpha(t)\rho(t) . \label{current-sop1}
\end{align}
Here we have introduced a current superoperator ${\cal
L}^{+}_\alpha(t)$ such that the current operator can be simply
expressed as the current superoperator acting on the reduced density
matrix. Note from Eq.~(\ref{uu-sol}) that $\sum_\alpha (\bm
\lambda^\beta_\alpha + \bm \lambda^{\beta\dag}_\alpha)=\bm
\gamma^\beta$ and $\sum_\alpha(\bm \kappa_\alpha + \bm
\kappa^\dag_\alpha)=2\bm \gamma$, the transient current operator
given by Eq.~(\ref{current-sop1}) is closely connected to the master
equation (\ref{emaster}) for the reduced density matrix $\rho(t)$.

Because of the trace over the states of the central system, the
transient current of Eq.~(\ref{tcd}) can be alternatively expressed
as
\begin{align}
I_\alpha(t)&=i \frac{e}{\hbar} \,{\rm tr}_{\rm s} \sum_i \big[a_iA^\dag_{\alpha i}(t)
 - A_{\alpha i}(t)a^\dag_i\big] \equiv - \frac{e}{\hbar}\,
 {\rm tr}_{\rm s}\hat{\tilde I}_\alpha(t).        \label{tcd-2}
\end{align}
Then using Eq.~(\ref{effao}) again, we have
\begin{align}
\hat{\tilde I}_\alpha(t)= & \sum_{ij} \Big\{ \bm
\lambda^\beta_{\alpha ij}(t)
[ a_j \rho(t)a^\dag_i +  \rho(t)a_ja^\dag_i] \notag \\
&+ \bm \kappa_{\alpha ij}(t)  a_j \rho(t)a^\dag_i + {\rm H.c.}
\Big\}\equiv {\cal L}^{-}_\alpha (t)\rho(t) , \label{current-sop2}
\end{align}
where ${\cal L}^{-}_\alpha (t)$ is defined as the superoperator for
$\hat{\tilde I}_\alpha(t)$. Now it is easy to check that the master equation (\ref{emaster})
for the reduced density matrix and the transient current of (\ref{curr0})
can be simply expressed as
\begin{subequations}
\label{DC}
\begin{align}
&\frac{d \rho(t)}{dt} = -i[H_{\rm S}(t), \rho(t)] + \sum_\alpha
[{\cal L}^{+}_\alpha(t)+{\cal L}^{-}_\alpha (t)]\rho(t) , \\
&I_\alpha (t) = \frac{e}{\hbar}{\rm tr_s}[{\cal L}^{+}_\alpha(t)\rho(t)] =
-\frac{e}{\hbar}{\rm tr_s}[{\cal L}^{-}_\alpha (t)\rho(t)] ,
\end{align}
\end{subequations}
where $H_{\rm S}$ is the original Hamiltonian of the central system.
This analytical operator relation between the reduced density matrix
and the transient current shows explicitly the intimate connection
between quantum decoherence and quantum transport in nonequilibrium
dynamics. The reservoir-induced non-Markovian relaxation and
dephasing in the transport processes are manifested through the
superoperators of (\ref{current-sop1}) and (\ref{current-sop2})
acting on the reduced density matrix $\rho(t)$. The time-dependent
parameters $\bm \lambda^\beta_\alpha(t)$ and $\bm \kappa_\alpha (t)$
appeared in the superoperators are given by Eq.~(\ref{uu-sol}) which
are determined by the dissipation-fluctuation
 integrodifferential equation (\ref{uv-eq}).
This completes the nonequilibrium theory for transient electron
dynamics in nanostructures we concerned.  It should be pointed out
that Eq.~(\ref{DC}) was formally expressed in terms of the real-time
diagrammatic expansion and the hierarchical
expansion.\cite{Kon9616820,Jin08234703} Here we have the analytical solution.

\subsection{Current conservation law}
\label{theoME} Now we should derive the current conservation law for
a self-consistent check.  Consider the
equation of motion for the operator $a^\dag_i a_j$ in the Heisenberg
picture,
\begin{align}
i\frac{d}{dt}a^\dag_i a_j = [a^\dag_i & a_j , H] = \sum_l
\epsilon_{jl}a^\dag_i a_l - \epsilon_{li}a^\dag_l a_j \notag \\
& + \sum_{\alpha k}(V_{j\alpha k}a^\dag_ic_{\alpha k} - V^*_{i\alpha
k}c^\dag_{\alpha k}a_j).
\end{align}
Taking the expectation value of the above equation with respect to
the state $\rho^H_{\rm tot}$ (the total density matrix in the
Heisenberg picture) and then transforming it into the Sch\"{o}rdinger
picture, we obtain
\begin{align}
i\frac{d\bm \rho^{(1)}(t)}{dt}=[\bm \epsilon(t), \bm \rho^{(1)}(t)]
+ i \sum_\alpha \bm I_{\alpha}(t) .\label{r1eom-1}
\end{align}
Here we have used the definition of the single-particle reduced
density matrix again, $\bm \rho^{(1)}_{ij}(t)\equiv{\rm tr}[a^\dag_j
a_i \rho_{\rm tot}(t)]={\rm tr}_{s}[a^\dag_j a_i \rho(t)]$, and also
introduced a current matrix $\bm I_{\alpha ji}(t)\equiv i {\rm tr}_s
[A^\dag_{\alpha i}(t)a_j -a^\dag_i A_{\alpha j}(t)].$
Equivalently, the transient current of Eq.~(\ref{tcd}) is simply
given by $I_\alpha (t) = \frac{e}{\hbar}{\rm Tr} \bm I_{\alpha}(t)$.
In other words, the equation of motion for $\bm \rho^{(1)}(t)$ is
directly related to the transient current.

On the other hand, the equation of motion for the single particle
reduced density matrix can also be obtained easily from the exact
master equation (\ref{emaster}). The result is
\begin{align}
\frac{d\bm \rho^{(1)}}{dt}&=\dot{\boldsymbol{u}}
\boldsymbol{u}^{-1} \bm \rho^{(1)}+\bm \rho^{(1)}(\bm
u^\dag)^{-1}\dot{\bm u}^\dag-\bm
\gamma^\beta \notag \\
&=-i[\bm \epsilon, \bm \rho^{(1)}] -(\bm \kappa \bm \rho^{(1)}+{\rm
H.c.}) -\bm \gamma^\beta. \label{r1eom-2}
\end{align}
Using the expression for $\bm \gamma^\beta(t)$ given by
(\ref{Gam-beta}), and comparing Eqs.~(\ref{r1eom-1}) with
(\ref{r1eom-2}), we have
\begin{align}
\bm I_{\alpha}(t)=& - \big\{\bm \lambda^\beta_\alpha(t) + \bm
 \kappa_\alpha(t)\bm \rho^{(1)}(t) + {\rm H.c.} \big\}.
 \end{align}
The above equation
reproduces exactly the transient current of Eq.~(\ref{curr0}). This
also provides a self-consistent check for the expression of
transient current derived directly from the influence functional
approach.

Taking the trace over the both sides of Eq.~(\ref{r1eom-1}) and also
noting the fact that $N(t)={\rm Tr}\bm \rho^{(1)}(t)$, the total
electron occupation in the central system, we have
\begin{align}
e\, \frac{d N(t)}{dt}= \sum_\alpha I_{\alpha}(t) \equiv - I_{\rm
dis}(t) ,\label{r1eom-3}
\end{align}
where $I_{\rm dis}(t)$ is defined as the transient displacement
current.\cite{Fra02195319} It tells that sum over the currents
flowing from all the leads into the central region equals to the
change of the electron occupation in the central region, as we
expected. In the steady-state limit $t\rightarrow \infty$,
$\dot{N}(t)={\rm Tr}\dot{\bm \rho}^{(1)}(t)=0$ so that $I_{\rm
dis}(t)=0$, as a consequence of current conservation.

\subsection{Solution to single particle reduced density matrix
and transient current}
Furthermore, we can rewrite Eq.~(\ref{Gam-beta}) as
\begin{align}
\frac{d \bm v^\beta}{dt}=\dot{\bm u}\bm u^{-1}\bm v^\beta +\bm
v^\beta (\bm u^\dag)^{-1}\dot{\bm u}^\dag-\bm \gamma^\beta . \notag
\end{align}
Comparing between Eq.~(\ref{r1eom-1}) for the single particle
reduced density matrix $\bm \rho^{(1)}(t)$ and the above equation
for $\bm{v}^\beta(t)$, it indicates that the solution of $\bm
\rho^{(1)}(t)$ is just $\bm {v}^\beta(t)$, apart from an initial
function. Explicitly, Eq.~(\ref{r1eom-1}) and the above equation
lead to
\begin{align}
\frac{d}{dt}(\bm \rho^{(1)}-\bm v^\beta)=\dot{\bm {u}} \bm {u}^{-1}
(\bm \rho^{(1)}-\bm v^\beta)+ (\bm \rho^{(1)}-\bm v^\beta) (\bm
u^\dag)^{-1}\dot{\bm u}^\dag. \notag
\end{align}
It is easy to find from the above equation the following
relationship between $\bm \rho^{(1)}(t)$ and $\bm {v}^\beta(t)$:
\be\label{rhot-one} \bm \rho^{(1)}(t)=\bm v^\beta(t)+\bm u(t)\bm
\rho^{(1)}(t_0)\bm u^\dag(t), \ee where $\bm \rho^{(1)}(t_0)$ is the
initial single particle reduced density matrix. Accordingly, the
transient current (\ref{curr0}) is reduced to
\begin{align}
I_\alpha(t) = - \frac{2e}{\hbar}\, {\rm Re}\int_{t_0}^{t}&
\!d\tau\,{\rm Tr} \Big\{ \bm g_{\alpha}(t,\tau)\bm v^\beta(\tau)
-\bm g^{\beta}_{\alpha}(t,\tau)\bar{\bm u}(\tau) \notag \\
& + \bm g_\alpha(t,\tau)\bm u(\tau)\bm \rho^{(1)}(t_0) \bm u^\dag(t)
\Big\}. \label{curr1}
\end{align}
The last term shows the explicit dependence on the initial single
particle reduced density matrix (including the initial electron
occupation in each level and the electron quantum coherence between
different levels in the central region). This term is an important
ingredient in the study of transient dynamics for practical
manipulation of a quantum device in real time and it will usually
vanish in the steady-state limit $t \rightarrow \infty$.

\subsection{Steady-state limit}

To make an explicit comparison with the steady-state current in
terms of the non-equilibrium Green functions and the
Landauer-B\"{u}ttiker formula used in the literature, we
introduce the spectral density of the lead $\alpha$: $ \bm
\Gamma_{\alpha ij}(\omega) =2\pi \sum_k V_{\alpha i k}V^*_{\alpha j
k}\delta(\omega-\epsilon_{\alpha k})$ and take a time-independent
bias voltage explicitly. Then the two-time correlation functions of
the $\alpha$-lead can be written as
\begin{subequations}
\label{corr-t}
\begin{align}
\bm g_{\alpha}(\tau-\tau')
 &= \int \frac{d\omega}{2\pi} \bm \Gamma_{\alpha}(\omega)e^{-i\omega (\tau-\tau')},
 \label{gt} \\
\bm g^{\beta}_{\alpha}(\tau-\tau')
 &= \int \frac{d\omega}{2\pi} f_\alpha(\omega)\bm \Gamma_{\alpha}(\omega)
 e^{-i\omega(\tau-\tau')}, \label{gbetat}
 \end{align}
 \end{subequations}
where $f_\alpha(\omega) =
1/(e^{\beta_\alpha(\omega-\mu_\alpha)}+1)$ is the Fermi
distribution function of the $\alpha$-lead at the initial time
$t_0$. Using the Laplace transformation, i.e.,
$f(z)=\int^\infty_{t_0} dt e^{-z(t-t_0)} f(t)$ with $z=-i\omega$,
we have
\begin{align}
 {\bm g}_\alpha(\omega) =\int \frac{d\omega'}{2\pi}\,\bm
\Gamma_\alpha(\omega')\frac{i}{\omega-\omega'+i0^+} =i\bm
\Sigma_\alpha^r(\omega) ,
\end{align}
i.e., ${\bm g}_\alpha(\omega)$ is the retarded self-energy induced
from the lead $\alpha$. The Laplace transformation of
Eq.~(\ref{ut-eq}) for $\bm u(t)$ gives
\begin{align}
\bm u(\omega)=\frac{i}{\bm \epsilon-\omega-\bm \Sigma^r(\omega)}
=i\bm G^{r}(\omega)  \label{uG}
\end{align}
where $\bm G^{r}(\omega)$ is just the retarded Green function, and
$\bm \Sigma^r (\omega)$ sums over $\bm \Sigma^r_\alpha (\omega)$ for
all $\alpha$. The advanced Green function is simply given by
$\bar{\bm u}(\omega)=-i\bm G^a(\omega)=\bm u^\dag(\omega)$.
Furthermore, for the time-independent Hamiltonian, the explicit
solution of Eq.~(\ref{vt-eq}) is
\begin{align}
\bm v^\beta(\tau)=\int^\tau_{t_0} d\tau_1 \int_{t_0}^{t }d\tau_2 ~
{\bm u}(\tau_1)\bm g^\beta(\tau_2-\tau_1) {\bm u}^\dag (\tau_2) .
\end{align} Taking the long time limit: $t\rightarrow \infty$,
its Laplace transformation gives
\begin{align}
\bm v^\beta(\omega)&={\bm u}(\omega)\bm g^\beta(\omega) {\bm u}^\dag
(\omega) \notag \\ &= -i{\bm G}^r(\omega)\bm \Sigma^<(\omega) {\bm
G}^a (\omega)=-i\bm G^<(\omega).
\end{align}
Here we have also used the relation $\bm g^\beta(\omega)=-i\bm
\Sigma^<(\omega)$. More explicit relations between $\bm u(t),
\bar{\bm u}(t)$ and $\bm v^\beta(t)$ with the retarded, advanced and
lesser Green functions in the real-time domain are presented in
Appendix.

Substituting above results into Eq.~(\ref{curr1}) in the steady-state limit
$t\rightarrow \infty$, we obtain the steady-state
single particle reduced density matrix and current:
\begin{subequations}
\begin{align}
\bm \rho^{(1)}_{\rm st}& =\int \frac{d\omega}{2\pi}\bm G^r(\omega)\Big[\sum_\alpha
\bm \Gamma_\alpha(\omega) f_\alpha(\omega)\Big]\bm G^a(\omega), \\
I_{\alpha,\rm st}& = \frac{2e}{\hbar}{\rm Im}\int \frac{d\omega}{2\pi}\,
{\rm Tr}\Big\{\bm \Gamma_{\alpha}(\omega)\big[\bm G^<(\omega)
   -f_\alpha(\omega) \bm G^a(\omega) \big] \Big\} .
\end{align}
\end{subequations}
This reproduces the steady-state current in terms of the
nonequilibrium Green functions in the frequency domain that has been
widely used. If we consider specifically a system coupled with left
(source) and right (drain) electrodes, i.e. $\alpha=L$ and $R$,
respectively, and also assume that the spectral densities for the
left and right leads have the same energy dependence: $\bm
\Gamma_L(\omega) = \lambda \bm \Gamma_R (\omega)$, where $\lambda$
is a constant. Then the net steady-state current flowing from the
left to the right lead is given by
\begin{align}
I_{\rm st} & = \frac{2e}{\hbar} \int \frac{d\omega}{2\pi}\big[f_L(\omega)
-f_R(\omega)\big] {\cal T}(\omega) , \notag \\
& {\cal T}(\omega) ={\rm Tr}\Big\{\frac{\bm \Gamma_L(\omega)\bm
\Gamma_R(\omega)} {\bm \Gamma_L(\omega)+\bm \Gamma_R(\omega)}{\rm
Im}[\bm G^a(\omega)]\Big\}.
\end{align}
This is the generalized Landauer-B\"{u}ttiker formula.\cite{Hau98}
Thus, we have simply recovered the nonequilibrium transport theory
at the steady-state limit. In fact, we can also reproduce the
transient transport theory in terms of the nonequilibrium Green
function technique, as shown in Appendix.

We now summarize the main results derived in this Section. We obtain
the general formula of the transient current $I_\alpha(t)$ through
the lead $\alpha$ which is given by Eq.~(\ref{curr0}) or
equivalently Eq.~(\ref{curr1}), accompanied with the exact master
equation for the reduced density matrix $\rho(t)$. This general
transient current is valid for arbitrary bias and gate voltage
pulses with arbitrary couplings between the central system and the
leads. We also establish explicitly the connection of the transient
current with the reduced density matrix, i.e. Eq.~(\ref{DC}),
through the superoperators ${\cal L}^{+}_\alpha(t)$ and ${\cal
L}^{-}_\alpha(t)$ determined by Eq.~(\ref{current-sop1}) and
(\ref{current-sop2}), which encompass all the back-reaction effects
associated with the non-Markovian dynamics of the central system
interacting with the lead $\alpha$. The current conservation law is
also explicitly derived. The steady-state current formula based on
the nonequilibrium Green function technique is reproduced as a long
time limit. These results allow us to explicitly analyze the
time-dependent quantum transport phenomena intimately entangling
with the electron quantum coherence and non-Markovian dynamics
through the reduced density matrix. The latter describes completely
the evolution of electron quantum coherence inside the
nanoelectronics device. Therefore, once one solves the equation of
motion (\ref{uv-eq}) for $\bm u(\tau), \bar{\bm u}(\tau)$ and $\bm
v^\beta(\tau)$, the full nonequilibrium dynamics of the
nanostructures can be analyzed explicitly.

\section{Analytical and Numerical Illustrations}
\label{thdynam}
\subsection{Time-independent bias voltage in the WBL:
an analytic solution for transient dynamics} \label{th_Vcons} For
practical applications, we take a simple system as an illustration:
a single dot containing only one spinless level that couples to the
left and right leads. The Hamiltonian of the dot is simply written
as $H=\epsilon a^\dg a$. This system only contains two states, the
empty state $|0\ra$ and the occupied state $|1\ra$. All the
corresponding matrixes (denoted by the bold symbols) in the above
formula are then reduced to a single function, such as $\bm
u(t)=u(t), \bm v^\beta(t) = v^\beta(t)$, and $\bm \rho^{(1)}(t)=\la
1|\rho|1\ra=\rho_{11}(t)=N(t)$. We will first consider a
time-independent bias voltage $V$. For simplicity, we also assume
that the tunneling couplings between the leads and the dot as well
as the densities of states for the leads are energy-independent. In
other words, the spectral density becomes a constant
$\Gamma_\alpha(\omega)= \Gamma_\alpha$. The non-local time
correlation functions is reduced to
\begin{subequations}
\label{corr-t0}
\begin{align}
g_{\alpha}(\tau-\tau') &= \Gamma_\alpha \delta(\tau-\tau'),
 \label{gt0} \\
g^{\beta}_{\alpha}(\tau-\tau')
 &=  \Gamma_{\alpha}\int \frac{d\omega}{2\pi} f_\alpha(\omega)
 e^{-i\omega(\tau-\tau')} . \label{gbetat0}
 \end{align}
 \end{subequations}
 This corresponds to the wide band limit (WBL) in the literature.

To be explicit, we take the initial time $t_0=0$ and let $e=\hbar=1$
in the following calculations. In the WBL, the solution of
\Eq{uv-eq} is \bsube
\begin{align}
u(\tau)& =\exp \Big\{- (i \epsilon+\frac{\Gamma}{2})\tau\Big\}~,
~\bar{u}(\tau)=u^\dag(t-\tau) \\
v^\beta(t)&= v^\beta_{\rm st}+\int \frac{d\omega}{2\pi}\,
 \frac{\Gamma_L f_L(\omega)+\Gamma_R f_R(\omega)}
{( \epsilon-\omega)^2+({\Gamma}/{2})^2}
\nl&\quad\quad\quad~~~ \times
\Big\{e^{-\Gamma t} -2e^{-\Gamma t/2}\cos[(\epsilon-\omega)t]
\Big\},
\end{align}
\esube where $\Gamma=\Gamma_L+\Gamma_R$ and $v^\beta_{\rm st}$ is
the solution of $v^\beta(t)$ at the steady-state limit:
\begin{align}
v^\beta_{\rm st}&=\int \frac{d\omega}{2\pi}\,
 \frac{\Gamma_Lf_L(\omega)+\Gamma_Rf_R(\omega)}
{( \epsilon-\omega)^2+({\Gamma}/{2})^2} =\rho^{(1)}_{\rm st}.
\end{align}
The electron occupation of the dot is calculated by
Eq.~(\ref{rhot-one}) as \be N(t)=\rho^{(1)}(t)=e^{-\Gamma
t}\rho^{(1)}(0)+v^\beta(t),  \label{occupt} \ee where
$N(0)=\rho^{(1)}(0)$ is the initial electron occupation of in the
dot. Obviously, in the steady limit, $N_{\rm st}=v^\beta_{\rm st}$.
This is not surprised since $\rho^{(1)}(t)$ and $v^\beta(t)$ obey
the same equation of motion that must lead to the same result in the
steady-state limit.

The transient current can then be analytically obtained:
\begin{align}
I_\alpha(t)
&=I_{\alpha,\rm st}-\Gamma_\alpha [N(t)-N_{\rm st}]
\nl&\quad-e^{-\frac{\Gamma t}{2}}\int \frac{d\omega}{2\pi}
\frac{\Gamma_\alpha f_\alpha(\omega)} {(\epsilon-\omega)^2+({\Gamma}/{2})^2}
\Big\{ \Gamma\cos[(\epsilon-\omega)t]
\nl&\quad\quad\quad\quad~~~~~~~~~~~~
-2(\epsilon-\omega)\sin[(\epsilon-\omega)t]
\Big\},
\label{It_WBL}
\end{align}
where the steady-state current is
\begin{align}
I_{\alpha,\rm st}&=\Gamma_\alpha\int \frac{d\omega}{2\pi}\,
 \frac{ \Gamma_L[f_\alpha(\omega)-f_L(\omega)]
 +\Gamma_R[f_\alpha(\omega)-f_R(\omega)]}
{( \epsilon-\omega)^2+({\Gamma}/{2})^2}.
\end{align}
Due to charge conservation, it is necessary to check
the displacement current which obeys the relation (\ref{r1eom-3}).
In the WBL,
\begin{align}
I_{\rm dis}(t) &= \Gamma e^{-\Gamma
t}N(0)- \int \frac{d\omega}{2\pi}\, \frac{\Gamma_L f_L(\omega)+
\Gamma_R f_R(\omega)} {(
\epsilon-\omega)^2+({\Gamma}/{2})^2} \nl&\quad\quad\quad\quad \times
\Big\{ -\Gamma e^{-\Gamma t} +{\Gamma}e^{-\Gamma t/2}\cos[(
\epsilon-\omega)t] \nl&\quad\quad\quad\quad +2(
\epsilon-\omega)e^{-\Gamma t/2}\sin[(\epsilon-\omega)t] \Big\} \notag \\
&=-\frac{d{N(t)}}{dt} .
\end{align}
At the steady-state limit, the displacement current $I_{\rm dis}
(t\rightarrow \infty) = 0$, as a consequence of current
conservation. It is also straightforward to calculate the net
current:
\begin{align}
I_{\rm net}(t)& = I_L(t)-I_R(t)
\nl&=I_{\rm st}
-(\Gamma_L-\Gamma_R) [N(t)-N_{\rm st}]
 \nl&\quad
-e^{-\frac{\Gamma t}{2}}\int \frac{d\omega}{2\pi}
\frac{\Gamma_L f_L(\omega)-\Gamma_R f_R(\omega)} {(
\epsilon-\omega)^2+({\Gamma}/{2})^2}
\nl&\quad
\times
\Big\{ \Gamma\cos[(\epsilon-\omega)t]
-2(\epsilon-\omega)\sin[(\epsilon-\omega)t]
\Big\}
\end{align}
where the stationary net current is
\bsube
\begin{align}
I_{\rm st}&= 2\Gamma_L\Gamma_R\int \frac{d\omega}{2\pi}\,
 \frac{ f_L(\omega) -f_R(\omega)}
{( \epsilon-\omega)^2+({\Gamma}/{2})^2} .
\end{align}
\esube

As we can see, once we solve the dissipation-fluctuation
integrodifferential equations, Eq.~(\ref{uv-eq}), a complete
time-dependence of all the physical quantities, such as the
electron occupations in the central system and the currents
flowing from each lead to the central region, can be obtained with
the explicit dependence of the time and the initial electron
occupation without ambiguity. From \Eq{It_WBL}, we see that the
initial current $I_{\alpha}(0)=-\Gamma_\alpha N(0)$ which depends
on the initial occupation of the dot. This result is consistent
with the electron occupation in the dot, Eq.~(\ref{occupt}). For
zero initial occupation, the initial current is zero. Note that
some of the above results are also obtained recently using the
nonequilibrium Green function technique \cite{Sch08235110}.

\subsection{Non-Markovian memory structure}
Realistically, the spectral density of the leads must depend on the
energy. Here we take the energy dependence as a Lorentzian-type
form:\cite{Mac06085324,Jin08234703,Tu08235311}
\begin{align}
\Gamma_{\alpha}(\omega)=\frac{\Gamma_{\alpha}W^2_\alpha}
{(\omega-\mu_\alpha)^2+W^2_\alpha},
\end{align}
where $\Gamma_{\alpha}$ describes the coupling strength and
$W_\alpha$ is the line width of the source (drain) reservoir with
$\alpha=L (R)$. Obviously the WBL,
$\Gamma_{\alpha}(\omega)=\Gamma_{\alpha}$, is achieved by simply
letting $W_\alpha \rightarrow \infty$. The lead correlation
functions with time-independent voltage can be parameterized as
\cite{Jin08234703} \bsube \label{corr-t1}
\begin{align}
g_{\alpha}(t-\tau)& =\frac{\Gamma_{\alpha}W_\alpha}{2}
e^{-\gamma_{\alpha 0} (t-\tau)}, \label{gt1}
\\
g^\beta_{\alpha}(t-\tau) &=\sum^M_{m=0} \eta_{\alpha
m}e^{-\gamma_{\alpha m} (t-\tau)}. \label{gbetat1}
\end{align}
\esube The first term in \Eq{gbetat1} with $m=0$ arises from the
pole of the spectral density function, with
\begin{align}
\eta_{\alpha 0}
=\frac{\Gamma_{\alpha}W_\alpha/2}{1+e^{-i\beta_\alpha W_\alpha}}~,
~~ \gamma_{\alpha 0}=W_\alpha+i\mu_\alpha.
\end{align}
The other terms with $m>0$ ($M\rightarrow\infty$ in principle)
arise from the Matsubara poles, where the relevant parameters are
explicitly given as
\bsube
\begin{align}
\eta_{\alpha m}&= \frac{i}{\beta_\alpha}\Gamma_{\alpha
}(-i\gamma_{\alpha m}), ~~m=1,\cdots \infty,
\\
\gamma_{\alpha m}&=\frac{(2m-1)\pi}{\beta_\alpha}+i\mu_\alpha.
\end{align}
\esube There are four typical timescales in an open system to
characterize the non-Markovian memory structure: the timescale of
the central system ($\sim 1/\epsilon$), the timescale of the
reservoirs ($\sim 1/W_\alpha$), the mutual timescale due to the
coupling between the system and the reservoir ($\sim
1/\Gamma_\alpha$), and the thermal timescale ($\sim \beta_\alpha$ or
$1/\mu_\alpha$). The timescale of the central system is usually
fixed when the system is set up. Then comparing with the character
time of the central system, a smaller finite line width $W_\alpha$,
a relatively large coupling $\Gamma_\alpha$, a low temperature
$1/\beta_\alpha$, and a comparable bias voltage $eV=\mu_L-\mu_R$ to
the energy levels of the central system will lead to a stronger
non-Markovian memory processes for a Lorentzian-type spectral
density, as we have demonstrated in [\onlinecite{Tu08235311}].

\begin{figure}
\includegraphics*[width=1\columnwidth,angle=0]{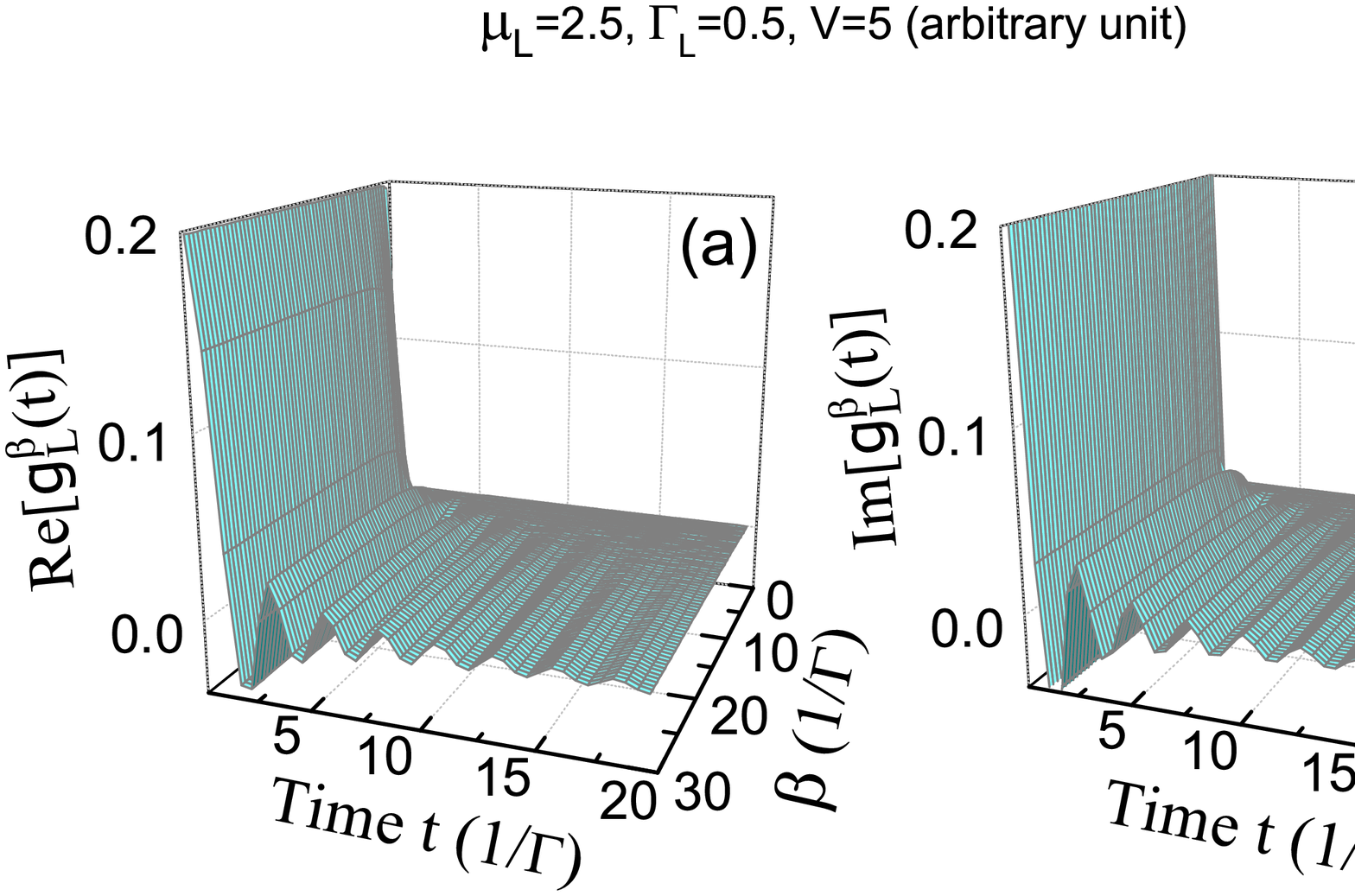}
\caption{(Color online) 3D plot of temperature-dependent non-local
time correlation function $g^\beta_L(t-\tau)$ (in unit:
$\Gamma^2$) as a function of temperature for $\mu_L=2.5\Gamma$ and
$\Gamma_L=0.5\Gamma$.}  \label{gbeta_beta}
\end{figure}

\begin{figure}
\centerline{\includegraphics*[width=1\columnwidth,angle=0]{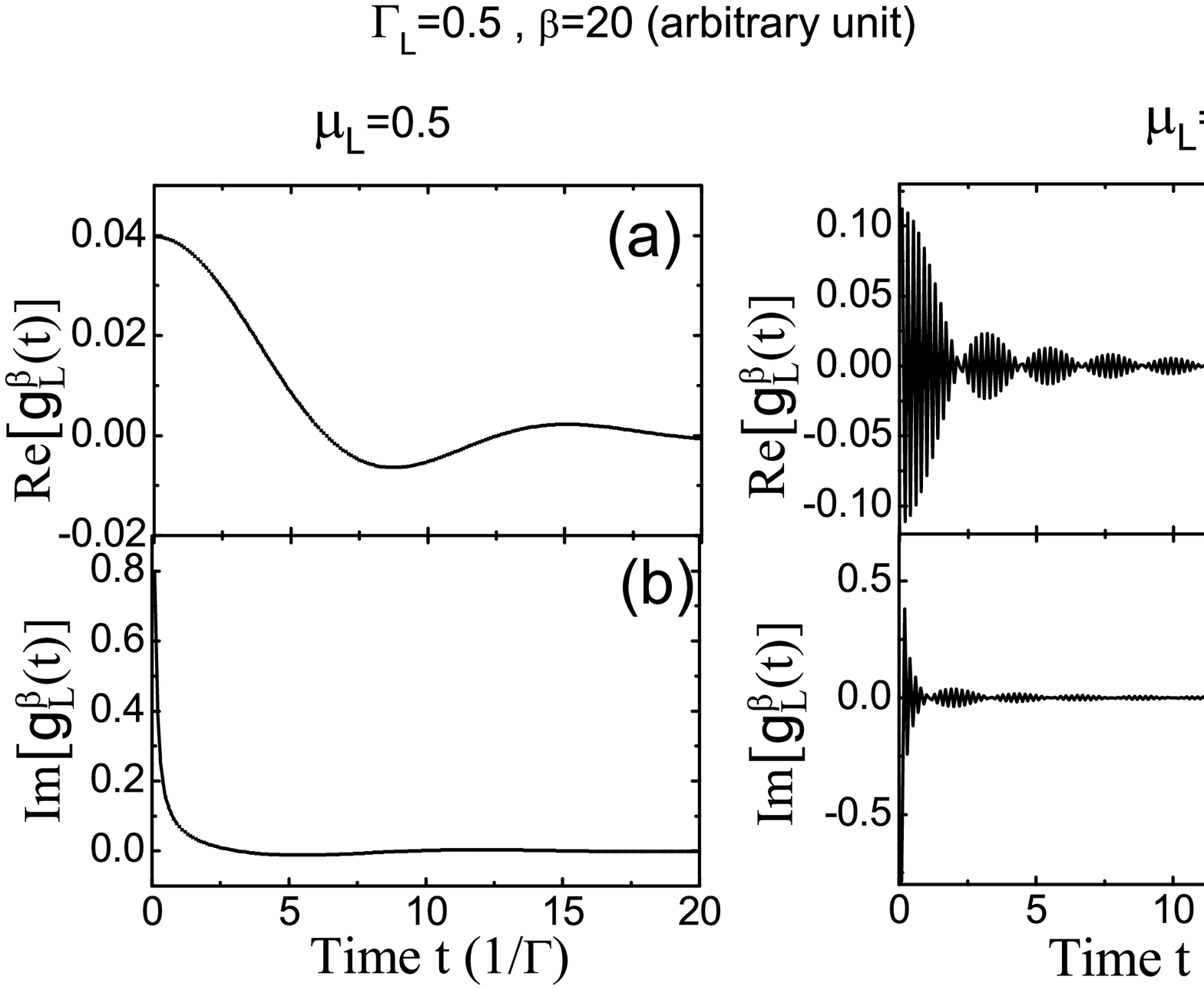}}
\caption{The temperature-dependent non-local time correlation
function $g^\beta_L(t-\tau)$ with different bias voltages at a
fixed temperature $\beta^{-1}=k_{\rm B}T=0.1\Gamma$ and
$\Gamma_L=0.5\Gamma$. The plots of (a) and (b) for
$\mu_L=0.5\Gamma$, (c) and (d) for $\mu_L=30\Gamma$ (the large
bias limit).}
 \label{gbeta_voltage}
\end{figure}

In fact, the line width $W_\alpha$ in a Lorentizan-type spectral
density is the main factor leading to the non-Markovian dynamics in
transient transport. In the WBL, $W_\alpha \rightarrow \infty$, the
dominated memory structure are mostly washed out. This can be seen
directly from the reservoir correlation functions. The correlation
function $g_{\alpha}(t-\tau)$ of \Eq{gt1} can be simplified to
$\frac{\Gamma_\alpha}{2}\delta(t-\tau)$ in the WBL. While for
$g^\beta_{\alpha}(t-\tau)$ in \Eq{gbetat1}, the first term ($m=0$)
is also simplified to a delta function of $t-\tau$ but the other
terms ($m\geq1$) are apparently changed not too much:
\begin{align}
 g^\beta_{\alpha}(t-\tau) \rightarrow
&\frac{\Gamma_{\alpha}}{2(1+e^{-i\beta_\alpha W_\alpha})}
\delta(t-\tau) \notag \\
& +\frac{i}{\beta_\alpha} \Gamma_{\alpha}\sum^M_{m=1}
e^{-\gamma_{\alpha m} (t-\tau)}. \label{gambeta2}
\end{align}
The profile of this temperature-dependent time correlation function
is plotted in Fig.\,\ref{gbeta_beta}. If we take further a high
temperature limit $\beta_\alpha\rightarrow 0$, the summation term in
\Eq{gambeta2} will also be reduced to a delta function of $t-\tau$,
see Fig.\,\ref{gbeta_beta}. Then no any memory effect remains, and a
true Markov limit is reached at high temperature limit. On the other
hand, for a large bias voltage limit $eV=\mu_L-\mu_R \rightarrow
\infty$, we have $f_L(\omega+\frac{eV}{2})\rightarrow1$ and
$f_R(\omega-\frac{eV}{2})\rightarrow0$ which lead to
$g^\beta_L(t-\tau)\rightarrow
g_L(t-\tau)=\frac{\Gamma_L}{2}\delta(t-\tau)$ and
$g^\beta_R(t-\tau)\rightarrow 0$ in the WBL. This reduces the
electron dynamics into a Markov limit again, as it has been widely
used in the literature.\cite{Gur9715215,Moz02161313,Fli04205334}
However, for a relatively low temperature or a finite bias voltage,
there appears to exist some non-Markovian effect in the WBL, coming
from the summation term in \Eq{gambeta2}. Fig.\,\ref{gbeta_voltage}
shows the time dependence of the correlation dependence with
different voltages. As one can see, the temperature-dependent
time-correlation function does not approach to a delta function in
time.

\begin{figure}
\centerline{\includegraphics*[width=1.0\columnwidth,angle=0]{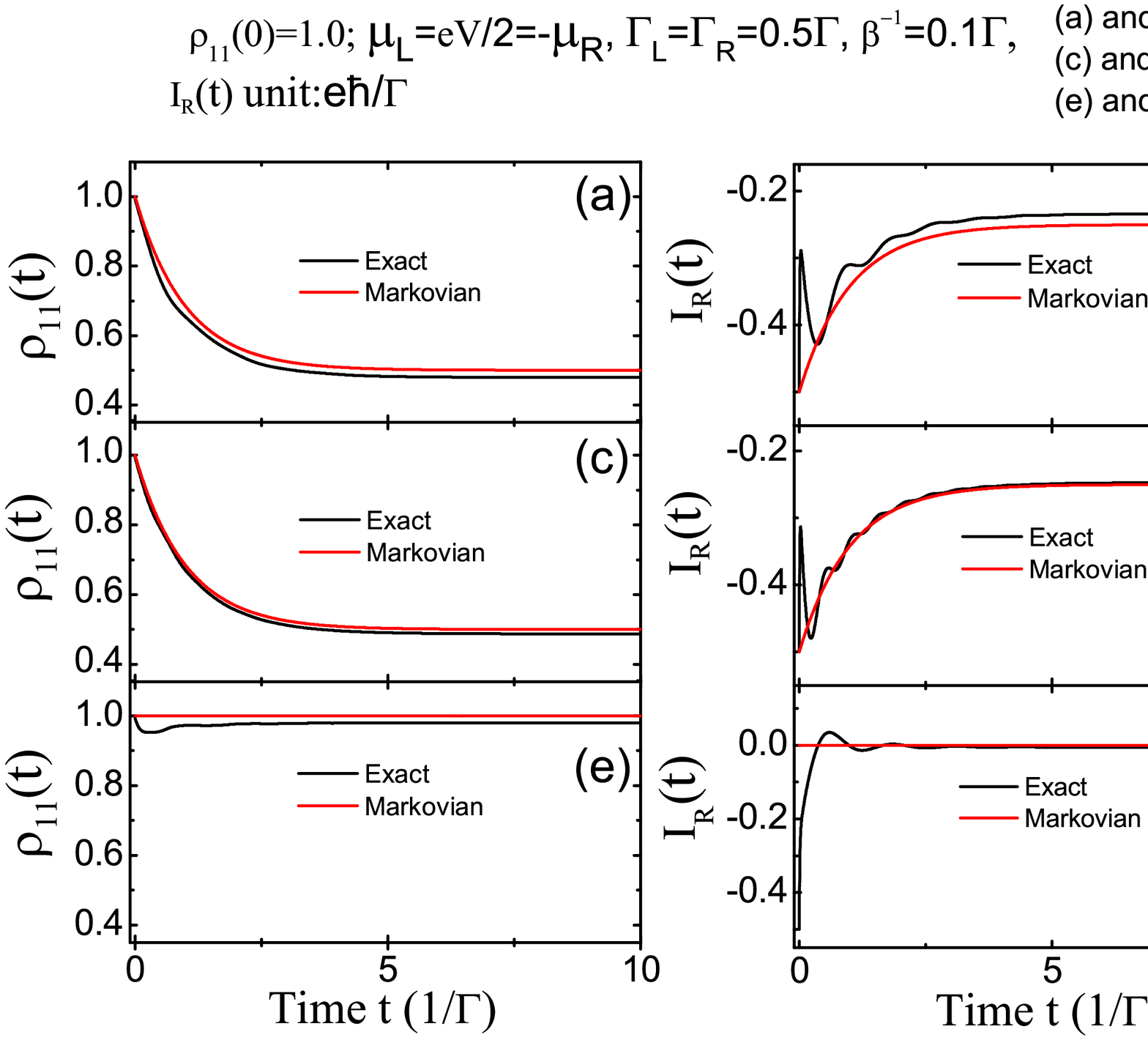}}
\caption{(Color online) Comparison of the non-Markovian dynamics
with Markov limit in the WBL for the electron occupation (left row)
and the transient current (right row). Here we take the parameters
$\Gamma_L=\Gamma_R=0.5\Gamma$, $\beta^{-1}= k_{\rm B}T=0.1\Gamma$,
$\mu_{L,R}=\pm eV/2=\pm 5 \Gamma$, and (a)-(b): $eV=10\Gamma$,
$\epsilon=2 \Gamma$; (c)-(d): $eV=20\Gamma$, $\epsilon=2 \Gamma$;
(e)-(f): $eV=10\Gamma$, $\epsilon=-10 \Gamma$. }
 \label{rho-curr-WBL}
\end{figure}

\begin{figure}
\centerline{\includegraphics*[width=0.9\columnwidth,angle=0]{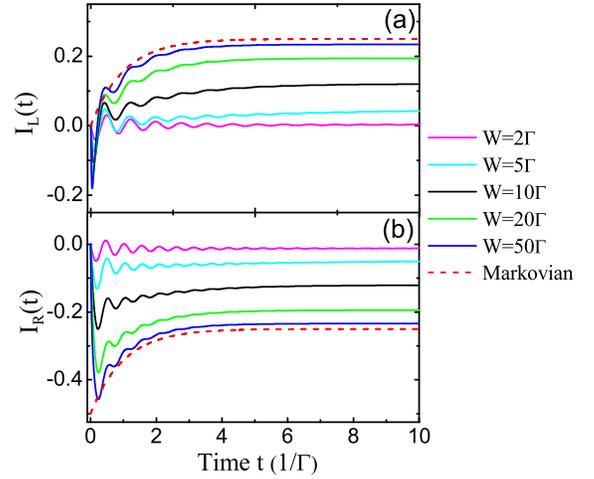}}
\caption{(Color online) The transient current $J_{L,R}(t)$ through
the left and right leads with a Lorentzian-type spectral density for
different bandwidths: dashed line, the WBL ($W=\infty$), other
lines, from the top to bottom for (a) and from the bottom to top for
(b), correspond to $W=50\Gamma, 20\Gamma, 10\Gamma, 5\Gamma,
2\Gamma$, respectively. Other parameters we used are
$\Gamma_L=\Gamma_R=0.5\Gamma$, $\beta^{-1}= k_{\rm B}T=0.1\Gamma$,
$\mu_{L,R}=\pm eV/2=\pm 5 \Gamma$, and $\epsilon=2 \Gamma$.}
 \label{curr-Widths}
\end{figure}

To examine if some non-Markovian effect can still survive in the
WBL, we numerically calculate the transient current passing through
the single level resonant tunneling nanostructure considered in the
last subsection. In the sequential tunneling regime
$\mu_L>\epsilon>\mu_R$, the exact solutions to the occupation and
the transient current are close to the Markov limit (differing by a
few present except for a very short time scale from the beginning),
as shown in Fig.\,\ref{rho-curr-WBL}(a)-(d). While, in the
co-tunneling regime with $\mu_\alpha\gg\epsilon$, the exact solution
of the occupation and the current are still almost the same as their
Markov solution except for the very short time scale from the
beginning, see Fig.\,\ref{rho-curr-WBL}(e)-(f). These results
indicate that the WBL (an extremely short character time of the
reservoirs) will dominatively suppress the thermal timescales. In
other words, when the line width $W \rightarrow \infty$, not only
for the high temperature and large bias limit, but even for a finite
temperature and a finite bias voltage, the non-Markovian effects
becomes quite weak that are most likely negligible in experiments.
Thus the WBL mainly takes into account the Markov dynamics. The
manifestation of the non-Markovian memory structure then should go
beyond the WBL.\cite{Tu08235311}  In Fig.\,\ref{curr-Widths}, we
calculate the transient current with different line widths to
demonstrate the non-Markovian effect in transport phenomena. As we
can see, the transport dynamics is significantly different from the
Markov limit for a small $W$. Increasing the value of $W$ will
decrease the memory effect accordingly. When the line width $W \ge
50\Gamma$, the exact solution of the transient current closely
approaches to the Markov result that is consistent with the WBL. A
analysis of the non-Markovian dynamics in this simple resonant
tunneling system has also recently been studied using Heisenberg
equations of motion.\cite{Sch09032110}


\subsection{Transient transport dynamics with time-dependent bias voltage}
\label{thcurr_Vt} We are now in the position to study the
transient transport phenomena in response to time-dependent bias
voltages. The calculation of the time-dependent electron current
in response to time-dependent bias voltages has received a great
attention recently, through various different theoretical
approaches.\cite{Jau945528,
Bru941076,Kon9616820,Sun004754,Zhu03224413,And05196801,
Mac06085324,Wei08195316,Sch08235110,Zhe09164708,Hau98} Transient
transport dynamics is also a central ingredient in many different
experiments, such as single-electron pumps and turnstiles with
time-modified gate signals moving electrons one by one through
quantum dots,\cite{Kir92,Swi991905,Blu07343,Fuj041323,Fuj08042102}
and the study of the quantum capacitance and inductances with ac
voltage
response.\cite{Nig06206804,Wan07155336,Zhe07195127,Zhe08184112,Mo09355301}
All these problems can be studied explicitly in the present theory
now. The corresponding nanoscale devices can be modelled as a
resonant tunneling nanostructure considered in this section. The
applying time-dependent voltages are taken to be the most commonly
interesting ones. For time-dependent voltages, as we mentioned in
Sec. II the single particle energy levels of the leads are changed
to $\epsilon_{\alpha k}(t)=\epsilon_{\alpha k}+eV_\alpha(t)$. The
non-local time correlation functions of the leads can be expressed
as \bsube \label{corr-vt1}
\begin{align}
 g_{\alpha }(\tau,\tau')
 &=\exp\left\{ -ie\int^\tau_{\tau'}d\tau_1 V_\alpha(\tau_1)\right\}
 g_{\alpha}(\tau-\tau'), \label{gvt1}
 \\
 g^{\beta }_{\alpha}(\tau,\tau')
 &=\exp\left\{ -ie\int^\tau_{\tau'}d\tau_1 V_\alpha(\tau_1)\right\}
 g^{\beta }_{\alpha}(\tau-\tau').
 \label{gbetavt1}
 \end{align}
\esube With the parametrization of Eq.~(\ref{corr-t1}), it is not
difficult to numerically calculate the transient electron dynamics
for arbitrary line width $W_\alpha$. In the following calculation,
we set the symmetric ac voltages, i.e., $\mu_L(t)=eV(t)/2$ and
$\mu_R(t)=-eV(t)/2$. For the quantum dot with a single level, the
reduced density matrix can be fully characterized by the electron
occupation in the dot. The exact numerical results for the
transient current and occupation due to different types of applied
ac voltages are presented as follows.

\subsubsection*{Exponentially time-dependent bias voltage}
\begin{figure}
\centerline{\includegraphics*[width=0.7\columnwidth,angle=0]{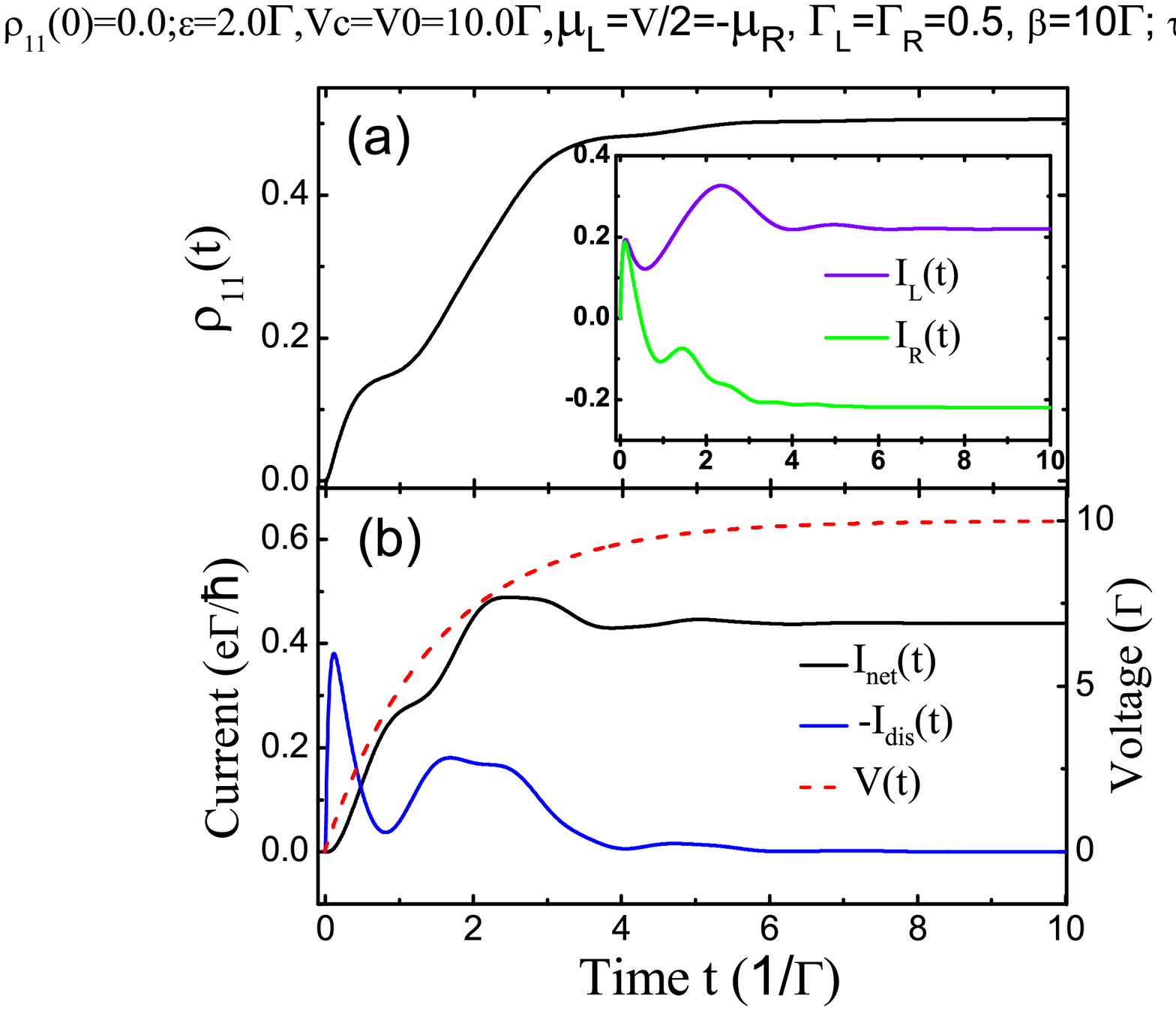}}
\caption{(Color online) The occupation and transient current
response to exponential time-dependent voltage  with
$\tau=1.5/\Gamma$. The other parameters are $eV=10\Gamma$,
$\Gamma_L=\Gamma_R=1.5\Gamma$, $\beta^{-1}=k_{\rm B}T=0.1\Gamma$,
$W_L=W_R=W=20\Gamma$ and $\epsilon=2\Gamma$.}
 \label{rho-curr-exp}
\end{figure}

We shall first study the transient transport dynamics in response
to an exponentially time-dependent bias voltage
$V(t)=V(1-e^{-t/\tau})$, where $\tau>0$ is a time dominating
switch-on rate of the voltage. The asymptotic limit
$\tau\rightarrow 0^+$ corresponds to a step function. The
numerical results are plotted in Fig.\,\ref{rho-curr-exp}.
The first peak shown in the displacement current arises from the
cotunneling process during the initially short time scale. At
beginning, the Fermi surface of the both leads are nearly equivalent
to zero and the dot energy level is higher than the Fermi surface.
The initial currents through both leads are equal to each other due
to the totally symmetric structure. This leads to a zero initial net
current. This feature is common for other type of ac bias voltages
discussing later. In general, the emergence of the peaks in the
current corresponds to a steep transient behavior of the electron
states (i.e. the occupation for the single-level dot) occurs inside
the dot. The non-linear response to the time-dependent bias is
clearly manifested in the change of both the electron state in the
dot and the transient current through the leads (including the
individual current through each lead and the displacement and net
currents), as we plotted in the Fig.\,\ref{rho-curr-exp}. In
particular, the net current changes in time that closely follows the
change of electron occupation in the dot, while the displacement
current depicts the steep change rate of the occupation, as we
expected from Eq.~(\ref{r1eom-3}).

\subsubsection*{Oscillating bias voltage} 
\begin{figure}
\centerline{\includegraphics*[width=0.7\columnwidth,angle=0]{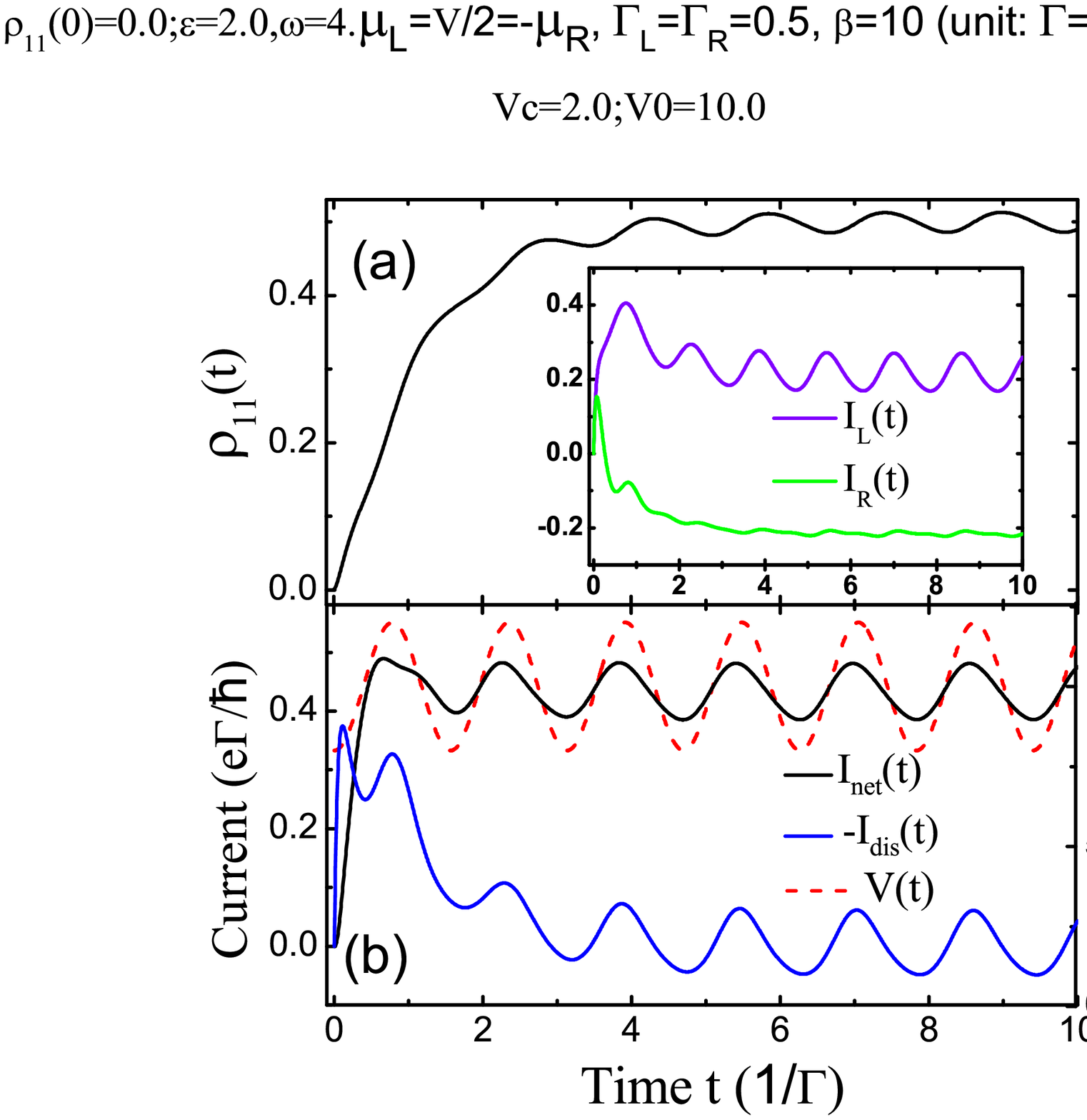}}
\caption{(Color online)  The occupation and transient current
response to oscillating voltage with the dc component
$eV_0=10\Gamma$, the ac component $eV_c=2\Gamma$ and the oscillating
frequency $\omega_c=4\Gamma$. The other parameters are the same as
in Fig.\,\ref{rho-curr-exp}.}
 \label{rho-curr-ac}
\end{figure}

\begin{figure}
\centerline{\includegraphics*[width=1\columnwidth,angle=0]{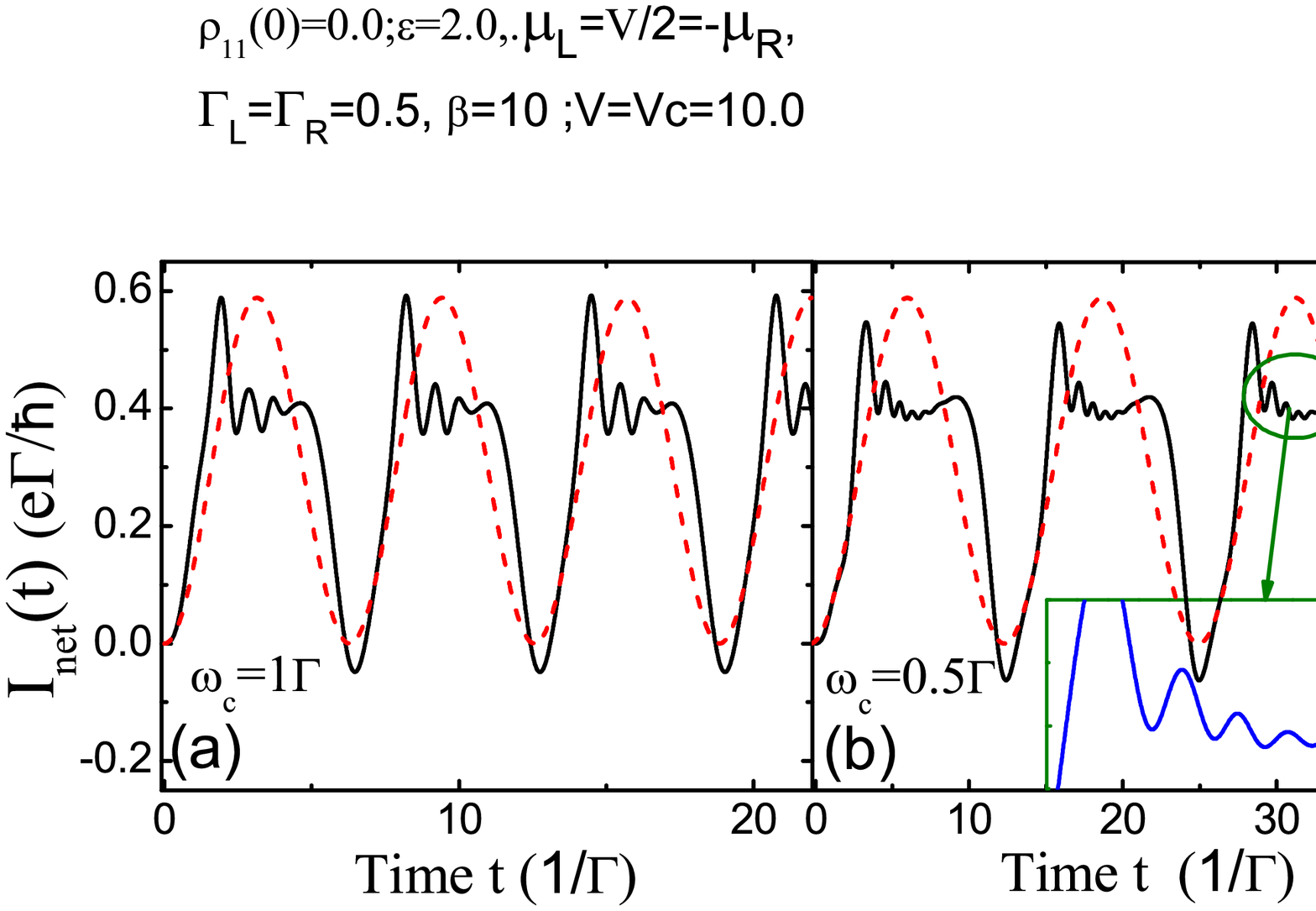}}
\caption{(Color online) The transient current (black line) response
to the oscillating voltage (the dashed red line) with
$eV_0=eV_c=10\Gamma$ and the oscillating frequency
$\omega_c=1\Gamma$ for (a) and $\omega_c=0.5\Gamma$ for (b).  The
other parameters are the same as in Fig.\,\ref{rho-curr-ac}.}
 \label{ac-net-Om}
\end{figure}

Now let us move to the
transient dynamics driven by an oscillating voltage:\cite{Hau98}
$V(t)=V_{0}-V_{c}\cos(\omega_c t)$, where $V_{0}$ is a dc
component, and $V_{c}$ and $\omega_c$ are the oscillation
amplitude and frequency of the ac component. The corresponding
exact numerical solution is shown in Fig.~\ref{rho-curr-ac}.
As one can see all the quantities, the electron state in the dot
and the transient currents, have the similar oscillation behavior
as the ac voltage oscillation, where the steady-state values and
the oscillation around the steady-state values are determined by
the dc voltage $V_0$. The oscillation amplitude around the
steady-state is proportional to the ac voltage amplitude $V_c$ and
the oscillation periodic is mainly given by $T=2\pi/\omega_c$.
However, the transient dynamics of the occupation and current does
not always strictly follow the ac voltage oscillation. The current
oscillation is a little slantwise compared to the ac voltage
oscillation. With increasing the amplitude of the ac voltage
$V_{c}$ and decreasing the oscillation frequency $\omega_c$, a
sideband oscillation occurs in the electron occupation as well as
in the transient currents as shown in Fig.\,\ref{ac-net-Om}.
Physically this sideband oscillation
is induced by the sinusoidal behavior of the ac voltage. It can be
understood analytically under the WBL:\cite{Hau98}
\begin{widetext}
\bsube
\begin{align}
&g^\beta_\alpha(\tau,\tau')=\sum_{n_1,n_2}J_{n_1}
\left({\Delta_\alpha\over\omega_c}\right)
J_{n_2}\left({\Delta_\alpha\over\omega_c}\right)\int
\frac{d\omega}{2\pi}\, \Gamma_\alpha f_\alpha(\omega)
 e^{-i(\omega+n_1\omega_c)\tau}e^{i(\omega+n_2\omega_c)\tau'},
\\
&v^\beta(t)=\sum_{\alpha,n_1,n_2}J_{n_1}
\left({\Delta_\alpha\over\omega_c}\right)
J_{n_2}\left({\Delta_\alpha\over\omega_c}\right)\int
\frac{d\omega}{2\pi}\, \Gamma_\alpha f_\alpha(\omega)
\frac{e^{-\Gamma t}+e^{-i(n_1-n_2)\omega_c t}-e^{-{\Gamma\over 2} t}
\left[e^{-i(\epsilon-\omega-n_2\omega_c)
t}+e^{i(\epsilon-\omega-n_1\omega_c) t}\right] }
{(\epsilon-\omega-n_1\omega_c)(\epsilon-\omega-n_2\omega_c)
+({\Gamma\over 2})^2-i{\Gamma\over 2} (n_1-n_2)\omega_c},
\\
&I_\alpha(t)=-\Gamma_\alpha\left\{(e^{-\Gamma t}N(t_0)+{\rm
Re}[v^\beta(t)]\right\}
+\sum_{n_1,n_2}J_{n_1}\left({\Delta_\alpha\over\omega_c}\right)
J_{n_2}\left({\Delta_\alpha\over\omega_c}\right)\int
\frac{d\omega}{2\pi}\, \Gamma_\alpha f_\alpha(\omega)2{\rm Im}
\frac{e^{-i(\epsilon-\omega-n_2\omega_c) t}-e^{-i(n_1-n_2)\omega_c
t}} {\epsilon-\omega-n_2\omega_c+i{\Gamma\over 2} }.
\end{align}
\esube
\end{widetext}
where the Bessel function satisfies $J_{-n}(z) = (-1)^nJ_n(z)$ and
$\Delta_{L,R}= \mp \frac{eV_c}{2}$.

\subsubsection*{Gaussian pulse}
\begin{figure}
\centerline{\includegraphics*[width=0.7\columnwidth,angle=0]{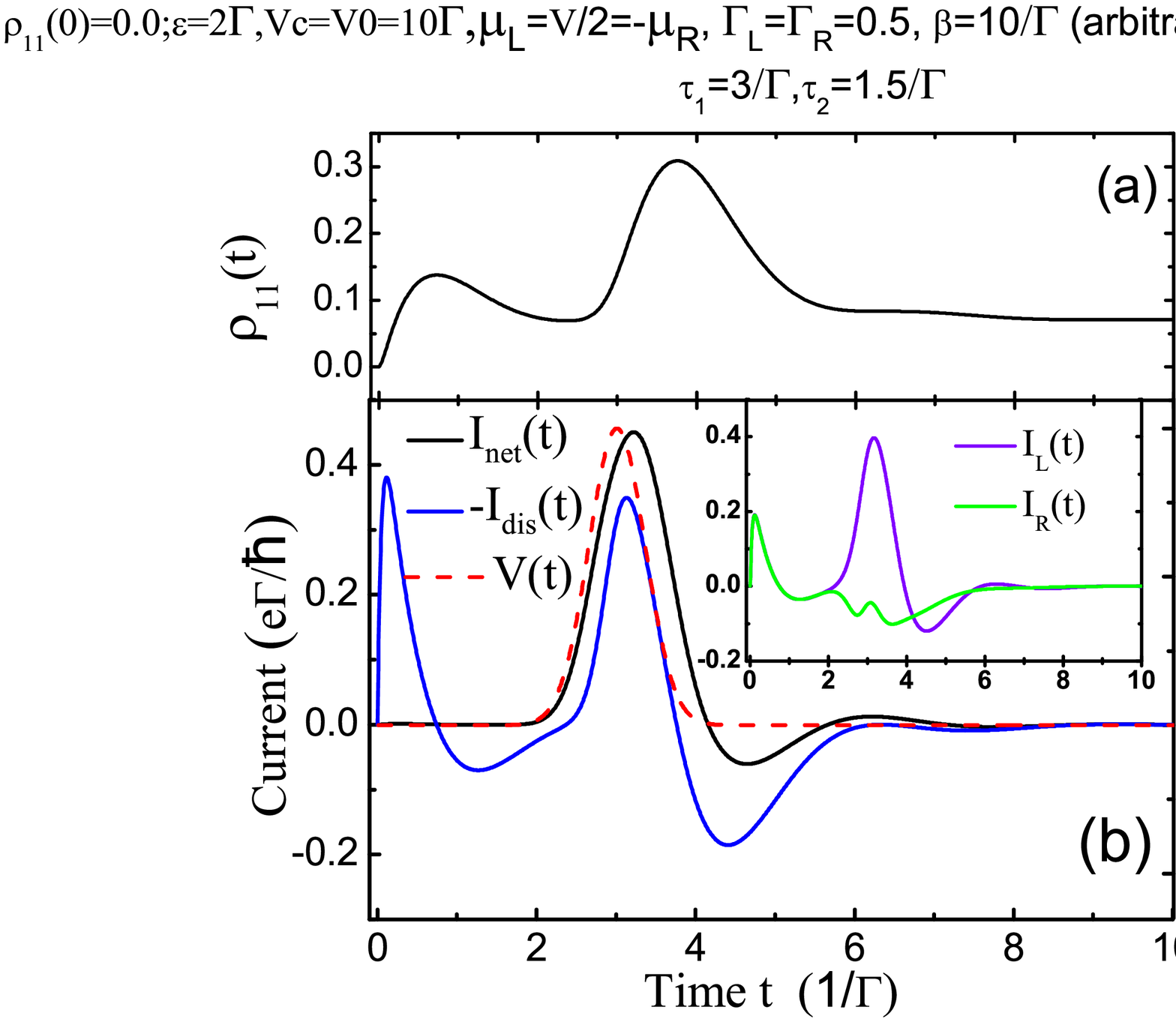}}
\caption{(Color online) The occupation and transient current
responses to Gaussian voltage pulse with $\tau_1=3/\Gamma$ and
$\tau_2=0.5/\Gamma$. The other parameters are the same as in
Fig.\,\ref{rho-curr-exp}.}
 \label{rho-curr-gauss}
\end{figure}

The last example we shall study is the transient dynamics droven
by a Gaussian pulse
$V(t)=V\exp\{-\frac{(t-\tau_1)^2}{\tau^2_2}\}$.  The width and the
center of the pulse are determined by $\tau_2$ and $\tau_1$,
respectively. The exact numerical result plotted in
Fig.\,\ref{rho-curr-gauss} shows that the net current peak emerges
(a slightly delay) just after the voltage pulse while the
corresponding response of the electron occupation delays
significantly. This behavior is easy to be understood that the
external voltage pulse leads to a sharp change of the electron
occupation due to the delay response. The shape change of the
transient current comes from the largest change rate of the
electron occupation in the dot. The delay response effect is
determined by the tunneling rate $\Gamma$. Outside the voltage
pulse, the tunneling current comes completely from the cotunneling
effect and the occupation in the dot decays to a stationary value.
It is interesting to note that the response of the occupation and
the currents to the cotunneling process is different before and
after the voltage pulse. Before the voltage pulse, the system is
dominated by the cotunneling process because the Fermi surface of
the both leads are nearly equivalent but is lower than the dot
energy level ($\mu_L \simeq \mu_R \simeq 0 < \epsilon=2\Gamma$). The
aligned fermi surfaces double the peak amplitude of the displacement
current which also gives the zero net current, while the occupation
has a corresponding change regarding the change of the displacement
current. With the voltage increasing, the Fermi surface of the left
lead moves up over the dot energy level while that of the right lead
moves down below the dot energy level such that the system gradually
approaches the sequential tunneling regime. This drives the electron
flowing from the left lead to the dot and then to the right lead.
Such a process results in a sensitive response of all physical
quantities, the electron occupation in the dot, the displacement and
the net currents.
Then with the voltage decaying to zero, the electron residing in the
dot is favorable to cotunnel to the left lead which gives a negative
current and finally reaches the steady-state. The transient electron
dynamics with the non-linear response to this Gaussian pulse is in
particular useful for quantum feedback control to the electron
states in the dot through the transient current that we will study
in the future. It is also favorable to study the quantum capacitance
and inductances.\cite{Zhe08184112,Mo09355301}

\begin{figure}
\centerline{\includegraphics*[width=0.7\columnwidth,angle=0]{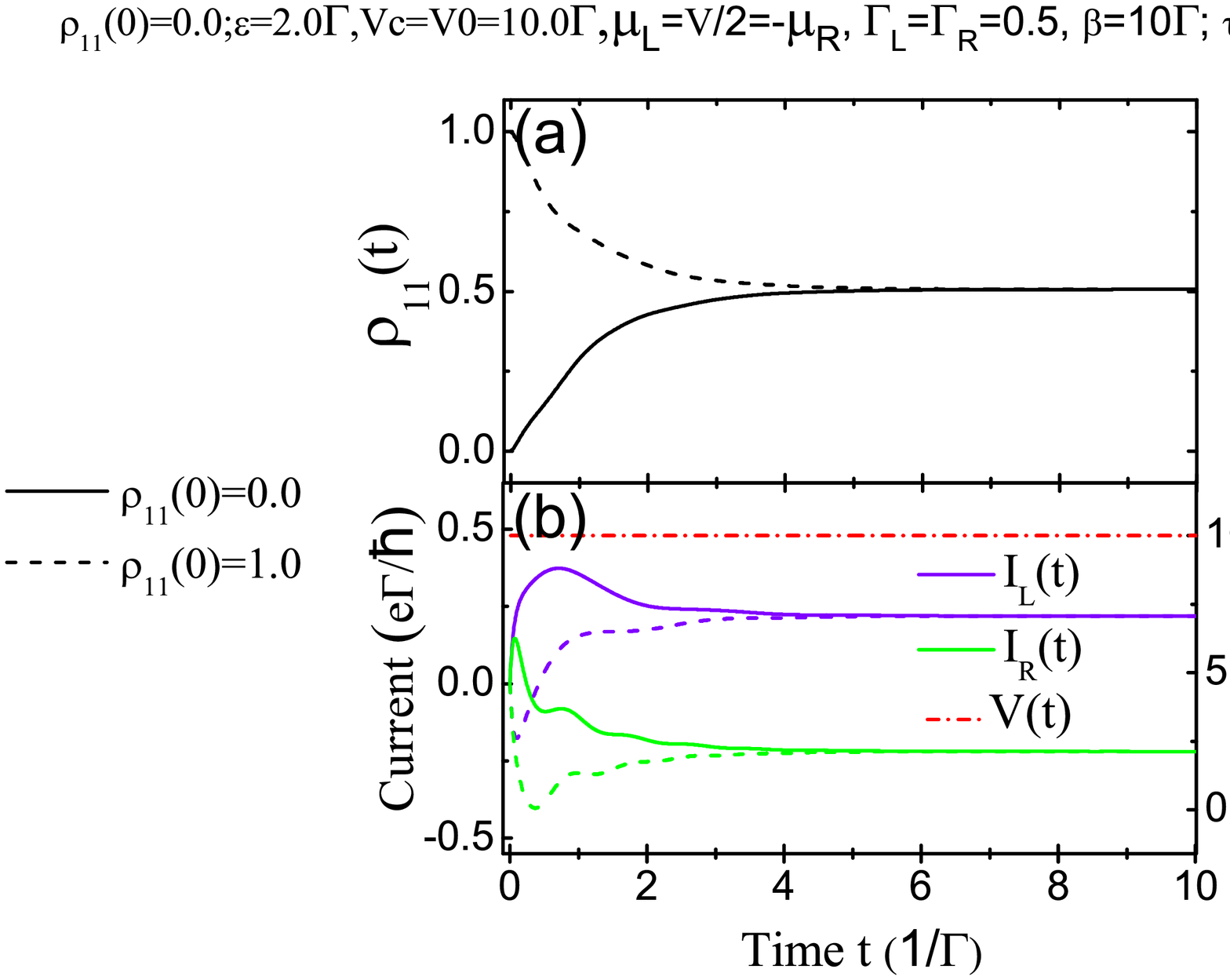}}
\caption{(Color online) The occupation and transient current
response to step pulse voltage (i.e. a constant bias after $t=0$,
the dot-dashed line) for the initial condition of
$\rho_{11}(t_0)=0$ (solid lines) and $\rho_{11}(t_0)=1.0$ (dashed
lines), respectively. The other parameters are the same as in
Fig.\,\ref{rho-curr-exp}.}
 \label{rho-curr-step}
\end{figure}

Note that the above transient dynamics starts with a zero initial
occupation, i.e., $\rho_{00}(t_0)=1$ and
$\rho_{11}(t_0)=1-\rho_{00}(t_0)=0$ [$N(t_0)=0$]. This implies
that the last term of \Eq{curr1} which is often ignored in the
nonequilibrium Green function technique\cite{Jau945528,Mac06085324}
has no contribution to the transient current in the above numerical
calculations. If the dot is initially occupied, the transient
current will be quite different although the steady-state limit is
the same because the initial electron distribution vanishes in the
steady-state limit. In Fig.\,\ref{rho-curr-step}, we show the exact
numerical result with such a situation where we take the simple
step-pulse voltage as an example. It shows that the initial
occupation in the central region has measurable contributions in
studying the transient dynamics, especially for the ultrafast
(extremely short time) operations in a quantum device.

Combining all these analysis together, as one can see the exact
numerical solutions presented here have demonstrated that both the
electron states in the dot and the transient currents past through
it have a clear non-linear response to the external bias voltage
pulses, in particular within a short time scale after the pulse is
turned on. These ultrafast non-linear response properties will
provide very useful information for the manipulation of the device
states as well as the quantum feedback controls for practical
applications. However, since we only consider here a very simple
device with a single-level dot, the quantum coherence and
decoherence dynamics to the transient current do not be manifested
in these numerical solutions. To demonstrate the quantum coherence
properties in the transient transport dynamics, it is necessary to
have the device containing at least two levels (or a single level
with electron spin degrees of freedom) in the central region.
Also, in the above numerical calculations, we take a rather large
finite line width, $W_\alpha=20 \Gamma$. The non-Markovian memory
structure will become more significant when the line width becomes
narrower.\cite{Mac06085324,Tu08235311} We will examine in detail
these features within the present theory in our future work.

\section{Summary and prospective}
\label{thsum} In summary, we have established a nonequilibrium
theory for the transient quantum transport dynamics via the
Feynman-Vernon influence functional approach. We obtain an
analytical relation between the master equation,
Eq.~(\ref{emaster}), and the transient current, Eq.~(\ref{curr0}),
determined by the same time-dependent (non-Markovian) coefficients.
In particular, the master equation and the transient current are
explicitly related to each other in terms of the superoperators
acting on the reduced density matrix of the central system, see
Eq.~(\ref{DC}). The back-reaction effect of the gating electrodes to
the central system is fully taken into account by the time-dependent
coefficients in the master equation (\ref{emaster}) through the
dissipation-fluctuation integrodifferential equation (\ref{uv-eq}).
The non-Markovian memory structure is non-perturbatively built into
the integral kernels in these equations of motion. The resulting
transient current, Eq.~(\ref{curr0}) or (\ref{curr1}), is rather
simple and it recovers the steady-state current in the
nonequilibirum Green function technique and the
Landauer-B\"{u}ttiker formula at the long time limit. This exact
nonequilibrium formalism should provide a very intuitive picture of
how the change of the electron quantum coherence in the central
system is intimately related with the electron tunneling processes
through the leads, and therefore non-linearly responses to the
corresponding external bias controls.
This theory is  applicable to study a variety of quantum transport
phenomena involving explicitly non-Markovian quantum decoherence
behaviors, in both stationary and transient scenarios, at arbitrary
 temperatures of the different contacts in the weak Coulomb interaction regime.

As a simple illustration, we apply the theory to a simple model of
electron tunneling though a single level quantum dot. We obtain
all analytical solutions in the wide band limit (WBL) where the
dependence of the initial electron occupation in the dot is
explicitly manifested. Taking a more realistic spectral density
with Lorentzian shape, we show that the Markov limit is a good
approximation in the WBL. The non-Markovian memory effect is
dominated by a finite line width of the spectral density. Under
the ac bias voltage pulses (including the step pulse, the Gaussian
pulse and the oscillation pulse), we have demonstrated the
ultrafast non-linear response of the electron occupation and
currents to the ac bias. We find that the current evolutions are
more sensitive to the energetic configuration of the dot than the
occupation evolution. This feature is very useful for quantum
feedback controls in quantum information processing. More
applications will be presented in the future works. These include
the decoherence transport dynamics in quantum dot devices such as
quantum dot Aharonov-Bohm inteferometers; the nonequilibrium
dynamics and the real time monitoring of spin polarization
processes in nanostrucutres; also the transient transport dynamics
in molecular electronics, as well as the application to
bio-electronics such as DNA junctions, etc.

In the end, we should also point out that the present theory is
developed without considering the electron-electron interaction in
the central region and therefore it is mainly valid in the weak
Coulomb interaction regime. It is not difficult to extend the
present theory to the strong Coulomb Blockage regime by properly
excluding the doubly occupied states in the central region of the
nanostructure, as we have shown explicitly in
[\onlinecite{Tu08235311}].  Although studying the above two extreme
limits, the extremely weak and the extremely strong Coulomb
interaction regimes together, could reveal a significant
understanding to various quantum transport phenomena, there is
increasing discussion for the intermediate Coulomb interaction
regime \cite{Cha08245412,Sch09033302} where the analysis of
transport dynamics should become much more complicated. In this
situation, the path integrals of Eqs.~(\ref{ppg}) and (\ref{AO-ppg})
may not be carries out exactly, and therefore it is not easy to find
an exact master equation and an exact expression of the transient
current. But the master equation, Eq.~(\ref{emaster}), and the
transient current, Eq.~(\ref{curr0}), can still serve as a good
approximation with respect to the saddle point approximation or loop
expansion,\cite{zhang921900} where the Coulomb interaction must be
included self-consistently in the dissipation-fluctuation
integrodifferential equation (\ref{uv-eq}). This approximate
treatment could provide a basic understanding to the quantum
transport phenomena in the intermediate Coulomb interaction regime.
The detailed extension of the present theory to the interacting
electron systems is in progress now. Nevertheless, the analytical
nonequilibrium theory we presented in this paper has the advantage
of combining the decoherence dynamics with transient transport
dynamics together to explore time-dependent physical phenomena and
may also provide a basic theory for quantum feedback controls in
various nanoelectronics devices.

\section*{Acknowledgement}

This work is supported by the National Science Council (NSC) of
ROC under Contract No.~NSC-96-2112-M-006-011-MY3, the National
Natural Science Foundation of China under Grants No.10904029, and
the Research Grants Council of Hong Kong (604709). We are also
grateful the National Center for Theoretical Science of NSC for
the support.

\appendix*
\section{The relations between $\bm u(t), \bar{\bm u}(\tau)$, $\bm
v^\beta(t)$ and the nonequilibrium Green functions} As we see both
the master equation (\ref{emaster}) and the transient current
(\ref{curr1}) are completely determined by the propagating matrices
of the stationary paths: $\bm u(\tau), \bar{\bm u}(\tau)$ and $\bm
v^\beta(\tau)$. Here we shall show that these propagating matrices
are directly related to the retarded, advanced and lesser Green
functions in the nonequilibrium Green function technique. In our
previous work,\cite{Tu08235311} the propagating matrices $\bm
u(\tau), \bar{\bm u}(\tau)$ and $\bm v^\beta(\tau)$ were introduced
to simplify the stationary path equations of motion (with the
convention $\bm x=\bm y\bm z$ for $x_i=\sum_j y_{ij} z_j$) as
follows: \bsube \label{uv-matrix}
\begin{align}
& \boldsymbol{\xi}(\tau)={\bm u}(\tau)\boldsymbol{\xi}(t_0)
+{\bm v}^\beta(\tau)[\boldsymbol{\xi}(t)+\boldsymbol{\xi}'(t)], \\
&\boldsymbol{\xi}(\tau)+\boldsymbol{\xi}'(\tau)= \bar{\bm
u}(\tau)[\boldsymbol{\xi}(t)+\boldsymbol{\xi}'(t)].
\end{align} \esube
 In fact, Eq.~(\ref{uv-matrix}) shows that $\bm u(\tau)$
is a propagating matrix of the forward stationary paths $\bm
\xi(\tau)$ starting at $t_0$, while $\bm v^\beta(\tau)$ mixes the
forward path $\bm \xi(\tau)$ and the backward path $\bm \xi'(\tau)$
started backwardly from $t$, and $\bar{\bm u}(\tau)$ is a backward
propagating matrix of the stationary paths. In fact, \Eq{uv-matrix}
shows that the transformation matrices $\bm u(\tau), \bar{\bm
u}(\tau)$ and $\bm v^\beta(\tau)$ should be defined more precisely
as $\bm u(\tau) \equiv {\sf u}(\tau,t_0)$, $\bar{\bm u}(\tau) \equiv
{\sf u}^\dag(t,\tau)$ and $\bm v^\beta(\tau) \equiv {\sf
v}^\beta(\tau, t)$: \bsube \label{uv-GF}
\begin{align}
& \boldsymbol{\xi}(\tau)={\sf u}(\tau,t_0)\boldsymbol{\xi}(t_0)
+{\sf v}^\beta(\tau,t)[\boldsymbol{\xi}(t)+\boldsymbol{\xi}'(t)], \\
&\boldsymbol{\xi}(\tau)+\boldsymbol{\xi}'(\tau)= {\sf u}^\dag(t,
\tau)[\boldsymbol{\xi}(t)+\boldsymbol{\xi}'(t)],
\end{align} \esube
where the Grassmannian variables $\bm \xi(\tau)$ and $\bm
\xi'(\tau)$ represent the forward and backward electron paths in the
functional path integrals. Then \Eq{uv-matrix} directly tells that
${\sf u}(\tau, t_0)$ describes the electron propagation (represented
by $\bm \xi(\tau)$ in the Grassmannian space) from the initial time
$t_0$ to the time $\tau$ so that it is just the retarded Green
function, namely,
\begin{align} \bm u(\tau) &= {\sf u}(\tau, t_0) =
i\bm G^r(\tau,t_0) \notag \\
&=\theta(\tau-t_0)\langle \{a_i(\tau),a^\dag_j(t_0)\} \rangle .
\label{ugreen}
\end{align}
Eq.~(\ref{uG}) is a justification for this relation. While, ${\sf
u}^\dag(t,\tau)$ describes the inverse propagation of the electron
[or the backward propagation represented by $\bm \xi'(\tau)$] from
the time $t$ to the time $\tau$ such that it is indeed the advanced
Green function:
\begin{align}
\bar{\bm
u}(\tau) = {\sf u}^\dag(t,\tau)=-i\bm G^a(\tau, t).
\end{align}
Meantime,
Eq.~(\ref{ut-eq}) indicates that the time correlation function of
the $\alpha$-reservoir:
\begin{align}
\bm g_{\alpha ij}  & (\tau_1,\tau_2) = i
\bm \Sigma_{\alpha ij}^r(\tau_1,\tau_2) \notag \\
&=\theta(\tau_1-\tau_2)\sum_{k}V_{\alpha k i}V^\ast_{\alpha k j}
\la\{ c_{\alpha k}(\tau_1), c^\dg_{\alpha k}(\tau_2)\}\ra_{\B},
\end{align} as a back-reaction effect of the
$\alpha$-lead to the central system, is the retarded self-energy.
These relations are also justified by
Eqs.~(\ref{ut-eq}-\ref{ubt-eq}).

The function ${\sf v}^\beta(\tau,t)$ describes the electron
propagation mixing the forward and backward paths so that it is
related to the lesser Green function defined by $\bm
G^<_{ij}(\tau,t)\equiv i\langle a^\dag_j(t) a_i(\tau)\rangle$ in the
nonequilibrium Green function formalism. In fact, it is not
difficult to find the explicit solution of Eq.~(\ref{vt-eq}):
\begin{align}
\bm v^\beta(\tau)={\sf v}^\beta(\tau,t)=&\int^\tau_{t_0} d\tau_1
\int_{t_0}^{t }d\tau_2 ~ {\sf u}(\tau, \tau_1)\bm g^\beta(\tau_1,
\tau_2) {\sf u}^\dag (t,\tau_2) \end{align} which in terms of Green
functions becomes
\begin{align} {\sf v}^\beta(\tau, t) =& -i \int^\tau_{t_0} d\tau_1
\int_{t_0}^{t }d\tau_2 ~ \bm G^r(\tau, \tau_1)\bm
\Sigma^<(\tau_1,\tau_2) \bm G^a(\tau_2,t) , \label{vtau}
\end{align}
where
\begin{align}
\bm \Sigma^<(\tau_1,\tau_2) = i\bm g^\beta(\tau_1,
\tau_2),
\end{align} is the lesser component of the self-energy.
On the other hand, the single-particle reduced density matrix is
related to the lesser Green function by $\bm \rho^{(1)}(t)=-i\bm
G^<(\tau,t)|_{\tau=t}$. From the relation of Eq.~(\ref{rhot-one}),
we find that $\bm v^\beta(\tau)$ can be expressed in terms of the
lesser Green function as follows:
\begin{align}
\bm G^<(\tau,t)=&i[ \bm u(\tau) \bm \rho^{(1)}(t_0)\bm u^\dag(t)+ {\bm v}^\beta(\tau)]\notag \\
 =& \bm G^r(\tau,t_0)\bm G^<(t_0,t_0)\bm G^a(t_0,t)\notag \\
 &+\int^\tau_{t_0} d\tau_1
\int_{t_0}^{t }d\tau_2 ~ \bm G^r(\tau, \tau_1)\bm
\Sigma^<(\tau_1,\tau_2) \bm G^a(\tau_2,t) .
\label{VR2}
\end{align}
This provides indeed a general solution of the lesser Green function
in the nonequilibrium Green function technique. In the previous
investigation of transient dynamics, the first term in
Eq.~(\ref{VR2}) that sensitively depends on the initial electron
distribution in the central regions are often
ignored.\cite{Jau945528,Hau98,Mac06085324} In fact, the second term
in Eq.~(\ref{VR2}) can only be identified as the lesser Green
function $\bm G^<(\tau, t)$ in the steady-state limit.

Using the above explicit relations between $\bm u(t), \bar{\bm
u}(\tau)$, $\bm v^\beta(t)$ and the nonequilibrium retarded,
advanced and lesser Green functions, we immediately obtain the
transient current of Eq.~(\ref{curr1}) in terms of the
nonequilibrium Green functions:
\begin{align}
I_\alpha(t)= -\frac{2e}{\hbar}{\rm Re}\int_{t_0}^{t}\!d\tau\,{\rm
Tr}\Big\{
   & \bm \Sigma^r_{\alpha}(t,\tau)\bm G^<(\tau,t) \notag \\
    & +\bm \Sigma^{<}_{\alpha}(t,\tau)\bm G^a(\tau,t)  \Big\}.
    \label{curr-G}
\end{align}
This current has exactly the same form obtained form the
nonequilibrium Green function technique.\cite{Jau945528,Hau98}
However, as we have pointed out in the practical applications, one
usually uses the steady-state lesser Green function, namely ignoring
the first term in Eq.~(\ref{VR2}). This term vanishes in the
steady-state limit so that it will not affect the steady-state
current. It can also be dropped if one takes the initial time $t_0
\rightarrow -\infty$ so that the central region is assumed to be in
an empty state initially. However, this term which explicitly
depends on the initial single particle reduced density matrix
(including the initial electron occupation in each level and the
electron quantum coherence between different levels in the central
region) is crucial for practical manipulation of a real quantum
device. Only in the wide band limit (WBL) where the non-local time
correlation function
\begin{align}
\bm g_\alpha(t,\tau)=i\bm \Sigma^r_\alpha (t,\tau)=\bm \Gamma_\alpha
\delta(t-\tau),
\end{align} the integral of the first term in
Eq.~(\ref{curr-G}) is reduced to the single particle reduced density
matrix: $\bm G^<(t,t)=i\bm \rho^{(1)}(t)$), which results in
\begin{align}
I_\alpha (t)=& -\frac{e}{\hbar} {\rm Tr}\Big[ \bm \Gamma_\alpha \bm
\rho^{(1)}(t) - \bm \Gamma_\alpha\int
\frac{d\omega}{\pi} f_\alpha(\omega) \notag \\
& \times \int^t_{t_0}d\tau{\rm Im}
\big\{e^{-i[\omega(t-\tau)+e\int_{\tau}^{t}d\tau' V_\alpha(\tau')]}
\bm G^a(\tau,t) \big\}\Big] .
\end{align}
Thus, the ignored initial occupation dependence is recovered in
the WBL.


\end{document}